\documentclass[galaxies,article,submit,moreauthors,pdftex,10pt,a4paper]{mdpi} 
\preto{\abstractkeywords}{\nolinenumbers}

\firstpage{1} 
\articlenumber{x}
\doinum{10.3390/------}
\pubvolume{xx}
\pubyear{2017}
\copyrightyear{2017}
\externaleditor{Academic Editor: name}
\history{Received: date; Accepted: date; Published: date}

\Title {Microvariability in BL Lac: Zooming into the Innermost Blazar Regions}

\Author{Gopal Bhatta $^{1,\dagger,\ddagger}$\orcidA{}, and James R. Webb $^{2,3}$*}

\AuthorNames{Firstname Lastname, Firstname Lastname and Firstname Lastname}

\address{%
$^{1}$ \quad Astronomical Observatory,  Jagiellonian University, ul. Orla 171, 30-244 Krak\'ow, Poland; gopalbhatta716@gmail.com\\
$^{2}$ \quad Department of Physics, Florida International University, Miami, FL 33199, USA; webbj@fiu.com\\
$^{3}$ \quad Southeastern Association for Research in Astronomy (SARA)}
\corres{Correspondence: gopalbhatta716@gmail.com; Tel.: +x-xxx-xxx-xxxx}

\secondnote{These authors contributed equally to this work.}

\abstract{ In this work, we present the results of our multi-band microvariability study of the famous blazar BL Lac.  We performed microvariablity observations of the source in the optical VRI bands for 4 nights in the year 2016.  We studied the intranight flux and spectral variability of the source in detail with  an objective  to  characterize microvariability in the blazars, a frequently observed phenomenon in blazars.   The results show that the source often  displays a fast flux variability with an amplitude as large as $\sim$ 0.2 magnitude within a few hours, and that the color variability in the similar time scales  can be characterized as  “bluer-when-brighter” trend. We also observed markedly curved optical spectrum during one of the nights. Furthermore, the correlation between multi-band emission shows that in general the emission in all the bands are highly correlated; and in one of the nights V band emission was found to lead the I band emission by $\sim$20 minutes. The search for characteristic timescale using  auto-correlation function and the structure function analyses  reveals characteristic timescale of $\sim$48 minutes in one of the R band observations. We try to explain the observed results in the context of the passage of shock waves through the relativistic outflows in blazars.} 

\keyword{Particles acceleration: non-thermal radiation; AGN: BL Lacertae objects: individual: BL Lac; galaxies: jets}

\begin{document}

\setcounter{section}{0} 

\section{Introduction}

Blazars, a subclass of radio galaxies with their jets pointing close to the earth, are known to exhibit some of the extreme properties such as high luminosity, rapid flux and polarization variability  and broadband non-thermal emission. These extreme properties are often attributed to the Doppler boosted  emission from the relativistic outflows emanating from the region close to the central engine \cite[e.g.,][]{Meier12}.   One of the most important properties that characterize blazars is their variability over a broad range of temporal and spatial frequencies. The variability in general is found to be aperiodic in nature; however, recently  detection of quasi-periodic oscillations in various emission frequencies and timescales have been reported\citep[see][]{Bhatta2017, Bhatta2016b,Zola2016}. In particular, low amplitude, rapid variability in the timescales less than a day is widely known as \emph{intraday/intranight} or  \emph{microvariability}. The terms microvariability and intranight usually refer to the  intraday variability as observed from the  ground based  optical telescopes. Following causality argument, the emission regions where microvariability is produced  should be highly compact (sub-)volumes, most likely, close to the central engine.  Such compact emission zones can not be spatially resolved  by any current instruments, therefore multifrequency microvariability studies could be one of the most powerful tools that allows us to zoom into the innermost regions at the interface of the supermssive black hole and the base of the blazar jets; and thereby constrain the nature of prevalent physical processes, e.g.  particle acceleration and energy dissipation mechanism, magnetic field geometry, jet content etc, that contribute to the origin of microvariability in radio-loud (RL) AGNs. To explain the phenomenon several  scenarios have been proposed most of which relate the source of the microvariability to a broad range of possible physical processes on-going both in the accretion disk and the jet. The scenarios include emission regions revolving around the central source, variable obscuration, various magnetohydrodynamic instabilities,  propagation of the shock waves down the turbulent jet, projection effects resulting from the orientation of the relativistic jets relative to the line of sight \cite[e.g.][]{MG85, Camenzind92, Wagner95,bhatta013}.  However, in spite of substantial observational efforts accompanied by comprehensive theoretical discussion on the subject, the details of the underlying processes responsible for microvariabity, and in general variability, still remain elusive.

Optical variability in blazars on intranight timescales  has been extensively studied over the past several decades \cite[for earlier review see][]{Wagner95}. In a sample of flat-spectrum radio sources  that have a compact Very Long Baseline Interferometry (VLBI) structure,  \cite{Quirrenbach1992} detected  intraday variations with amplitudes ranging up to 25 \%. Similarly,  \cite{Heidt1996} presented the results of the microvariability study in a large sample of radio-selected BL Lac objects which reported  the detection of microvariability in the sources with a large ($\sim 0.8$) duty cycle (DC) and  a typical peak-to-peak variability amplitude as large as $\sim$ 30\%. 
 In a study of several BL Lac objects consisting of the radio-selected BL Lac objects (RBLs) and X-ray-selected BL Lac objects (XBLs),   \cite{Bai1998} found that XBLs show more frequent microvariability than RBLs. 
 \cite{Kraus2003} observed that intraday variations are common in RL AGNs and that the flux variations  are often accompanied by similar variations in the linear polarization.   Similarly, during the study of long-term light curves of four blazars, including the source,  \cite{Howard2004}  found that the occurrences of microvariability were  temporally correlated with long-term optical activity of the source. In other words, microvariability is correlated with the flux gradients rather than  specific flux states. \cite{Rani2011}  reported the detection of microvariability with large DC  in  a sample of bright low-energy-peaked BL Lacs (LBLs).  Similarly, during their study of multiband optical flux and color variations of blazars on intraday and short term timescales of a few months \cite{Gaur2012} found that the BL Lac spectra often showed bluer-when-brighter trend whereas in flat spectrum radio quasars (FSRQ) the  redder-when-brighter trend appears more frequent. 
\cite{Bachev2012} found evidences of quasi-periodic oscillations at very low-amplitude levels during the short-term optical monitoring program of 13 blazars. Furthermore,  symmetry analysis of the longterm optical light curves of the source resulted a positive correlation between flare durations and peak fluxes \citep{Guo2016}. As an attempt to characterize  microvariability in the famous BL Lac S5 0716+714 robustly, the source was intensely studied using extended whole earth blazar telescope (WEBT) campaigns  \citep[see][]{bhatta013,bhatta015,bhatta016,Bhatta2016a}.
 More recently, in the hard X-ray regime the properties of intraday variability of a sample of radio-loud AGNs  are discussed by \cite{Bhatta2017b}.

BL Lacertae (RA=22h\,02m\,43.3s, Dec= +42d\,16m\,40s, and z = 0.0686), the eponymous blazar originally mistaken for a star, is one of the widely studied blazars across a wide range of electromagnetic frequencies using most of the currently available instruments. Because of its highly pronounced optical variability, it has been one of the most favorable targets for a large number of multifrequency studies  e.g. \citep{Wehrle2016,Raiteri2013,Raiteri2010, Villata2009,Bottcher2003}. On the intranight timescales, rapid flux and color variations with a trend to become bluer when brighter have been reported by several authors \cite[e.g.][]{Speziali1998,Massaro1998,Nesci1998,Clements2001,Zhai2012,Agarwal2015,Gaur2015}.
  The optical power spectral density (PSD) in the intranight timescale can be characterized as red noise behavior with a slope index of 2 \cite{Papadakis2003}. Cross-correlation between  longterm optical light curve and radio hardness ratio
revealed a radio time delay of more than 3 months \cite[see][]{Villata2004}. During  the multifrequency (radio to $\gamma$-rays) coordinated campaign by \cite{Abdo2011}, the source was found to be variable in $\gamma$-rays; and the Very Long Baseline Array (VLBA) observations revealed a synchrotron spectrum self-absorption turnover which allowed the authors to  constrain the average magnetic field of the regions to be less than 3 Gauss.  Similarly,  VLBA images also showed  superluminal features with the apparent speeds  ranging from 3.9 -- 13.5c \citep{Cohen2015}.  In addition, \cite{Marscher2008}, using radio and the optical polarization observation, reported the detection of a bright feature in the jet triggering a double flare in optical to TeV  energy ranges followed a delayed outburst at radio wavelengths. In the active state of the source during 2012-2013 multi-frequency (radio to  $\gamma$-rays ) observation campaign, \citep{Wehrle2016} noted time lags between the emission in bands, and the changing shapes of the spectral energy distributions (SED).    
The  Very Energetic Radiation Imaging Telescope Array System (VERITAS) observations of the source revealed one of the most rapid TeV $\gamma$-ray flares with  decay timescale of $\sim$13  minutes accompanied by the emergence of a  superluminal component from the VLBA core  as well as changes in the optical polarization angle \cite{Arlen2013}.
During an active phase of the source, \citep{Gaur2015} observed significant cross-correlations between optical and radio bands with a $\sim250$ days delay of cm-fluxes.  With a study of   photo-polarimetric observations  of the source \citep{Gaur2014} reported anti-correlation between the flux and polarization. 

 \begin{figure}[t!]
\centering
\begin{tabular}{c@{}c}
\resizebox{0.48\textwidth}{!}{\includegraphics{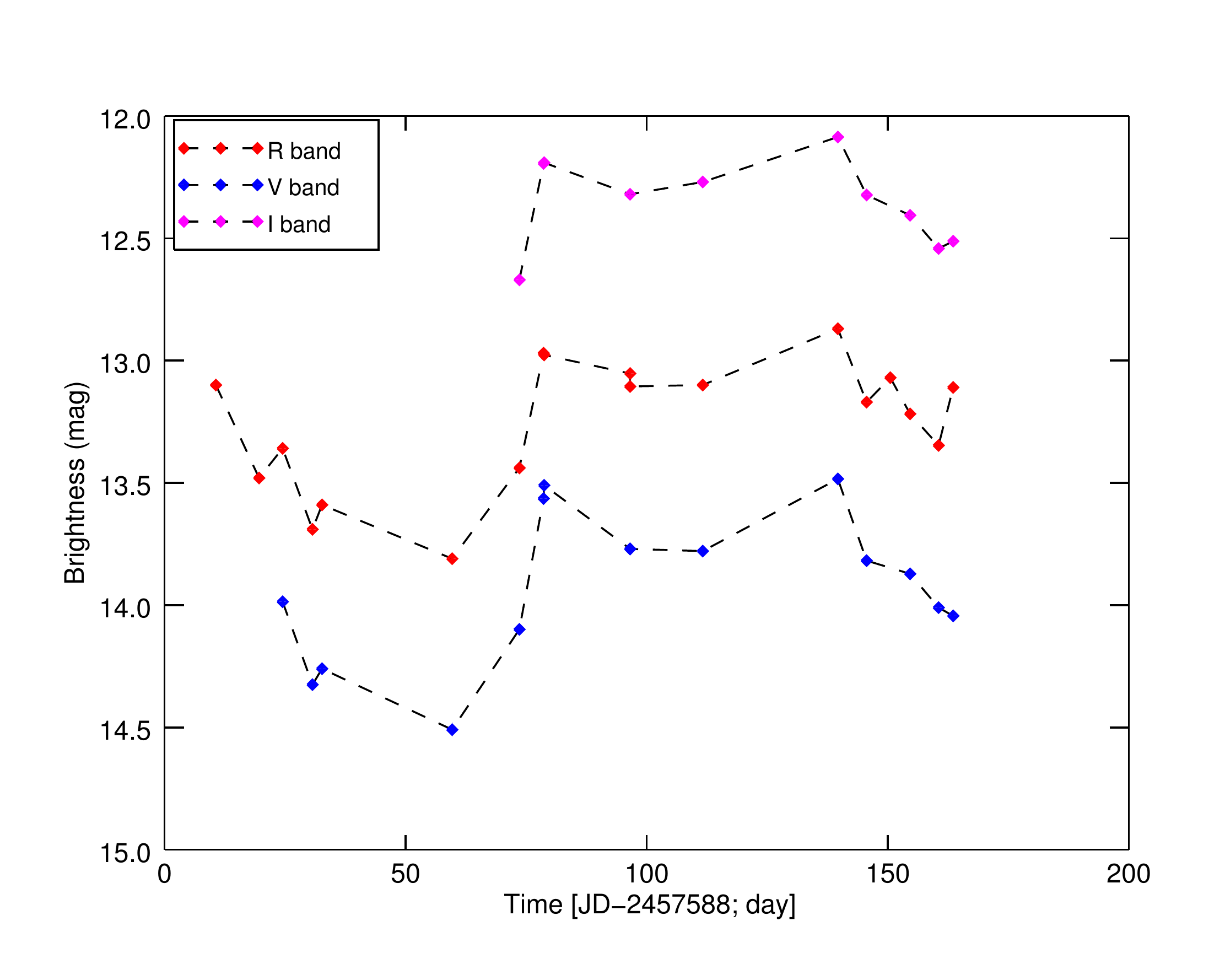}} &
\resizebox{0.48\textwidth}{!}{\includegraphics{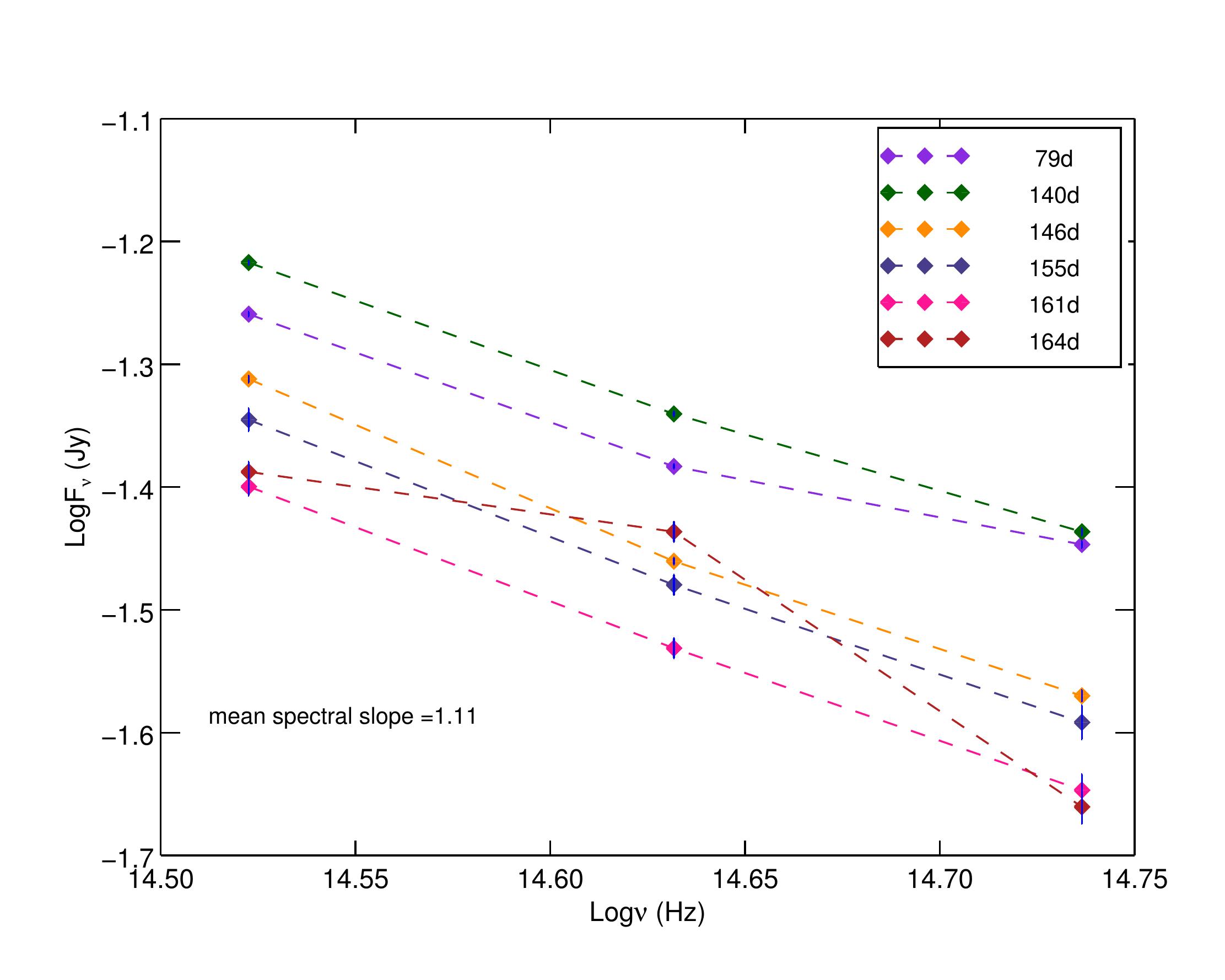}} \\
\end{tabular}
\caption{Left panel: Multi-band optical  monitoring of the blazar BL Lac in the year 2016. Right panel: The optical spectra of the source with the mean spectral index of 1.1}
\label{fig:1}
\end{figure}

In this paper, we present the results of our microvariability study of the blazar BL Lac in VRI photometric bands during 4 nights in 2016. The work represents a part of long-term and broader effort to characterize the nature of microvariablity in a sample of radio-loud AGNs  \citep[for detail see][]{Webb2016}.  We organize our presentations in the following way: In Section 2, the optical observations and the data processing are discussed. In Section 3, we elaborate on our flux and spectral variability analysis that includes cross-correlation study on the light curves along with auto-correlation and structure function analyses. Finally, we present our discussions and conclusions in Section 4.
 
\section{Observations and Data Processing}
\label{sec:obs}
Most of the microvariablity observations utilized in the work were acquired with the  Southeastern  Association for Research in Astronomy (SARA) suite of telescopes which includes the SARA 0.9-meter at Kitt Peak and the SARA Jacobus Kapteyn Telescope (JKT) at La Palma Canary islands. To extract the source brightness in magnitudes along with related uncertainties, the images were reduced and processed using the software MIRA\footnote{http://www.mirametrics.com/mira$\_$al.php}. The standard procedures for aperture photometry  were followed after the images were corrected for bias, dark, and flat-fielding. Apertures of about 2-4 arcseconds were chosen so as to minimize the flux scatter in the comparison stars in the same CCD frame. The comparison stars were selected from the Heidelberg AGN finder chart\footnote{ https://www.lsw.uni-heidelberg.de/projects/extragalactic/charts/2200+420.html}.   Star B in the field was excluded as  we suspected it to be slightly variable.  To determine if the microvariability is real or instrumental in nature we also routinely monitor the flux of the comparison stars; and  to minimize the sky effects  microvariability observations are performed on  the  nights with stable photometric conditions such as \emph{seeing}. The test for microvariability was performed  following the well-known statistical method discussed in \cite[][]{Howell1988}. 

We have been monitoring the source since past several years. However in this paper, we particularly discuss its variability properties in the year 2016. We observed the source continuously for a few hours in the nights dated 2016-08-07, 2016-08-12, 2016-09-30 and 2016-10-05 (hereafter Night 1, Night 2, Night 3 and Night 4, respectively) for micro-variability study.
 The source brightness in magnitudes was converted into the flux in mJy units by using the zero points for UBVRI-JHK Cousins-Glass-Johnsons system given in Table A2 of \citep{bessel98}, and to calculate the optical spectra the fluxes were interstellar-extinction corrected using the extinction magnitudes for various filters listed in the NED\footnote{www.ned.ipac.caltech.edu}.

  \begin{figure}[t!]
\centering
\begin{center}
\begin{tabular}{c@{}c}
\resizebox{0.48\textwidth}{!}{\includegraphics{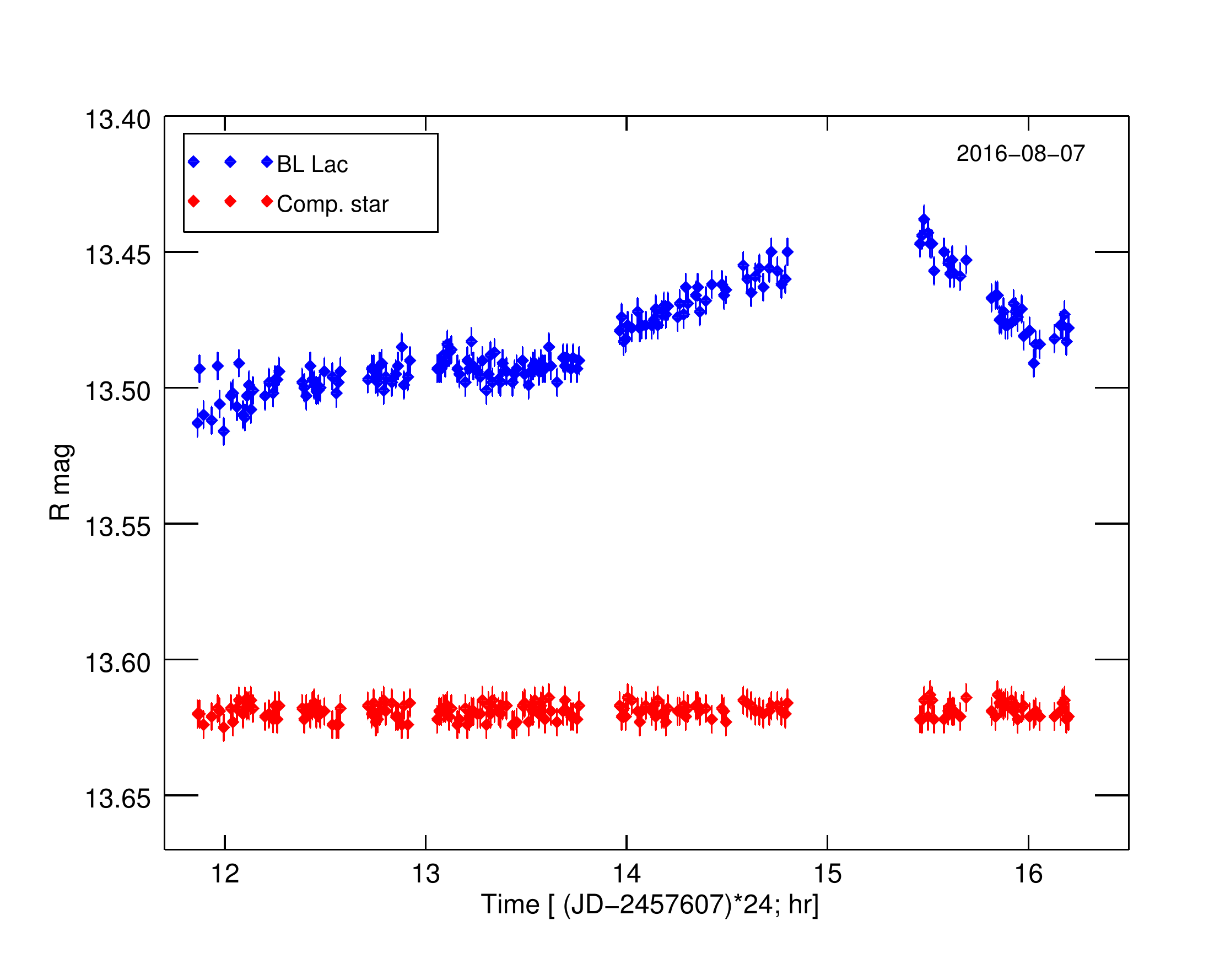}} &
\resizebox{0.48\textwidth}{!}{\includegraphics{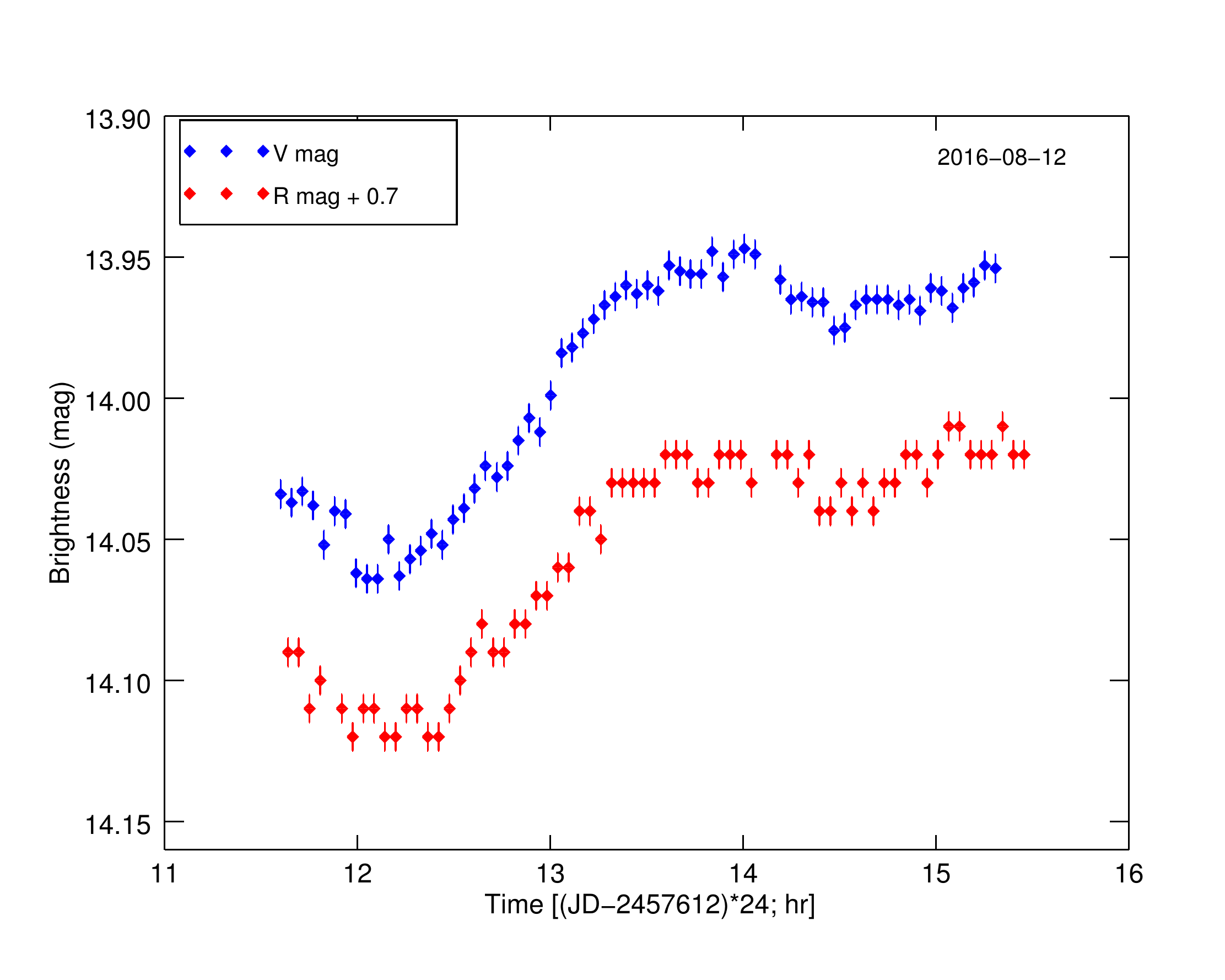}} \\
\resizebox{0.48\textwidth}{!}{\includegraphics{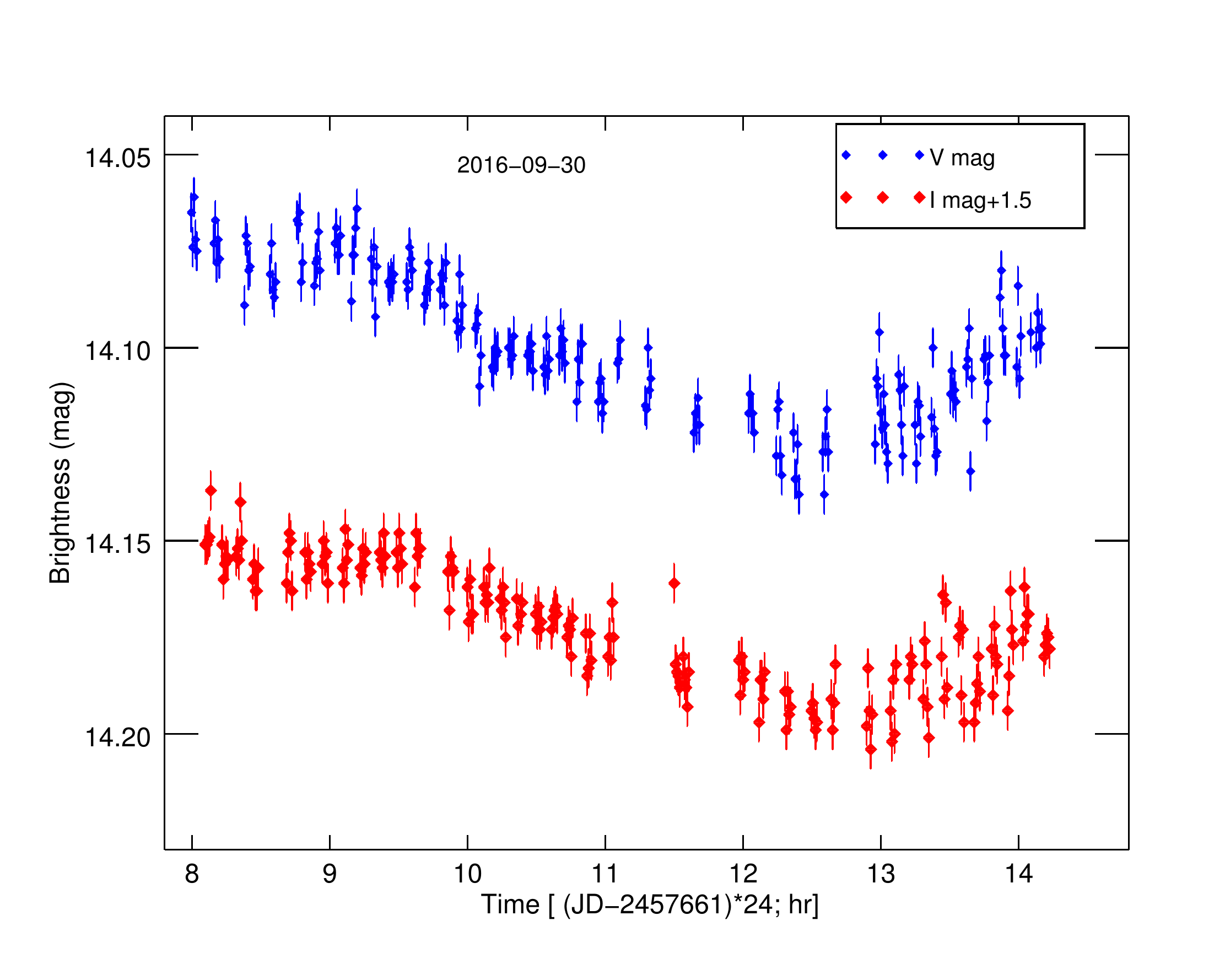}} &
\resizebox{0.48\textwidth}{!}{\includegraphics{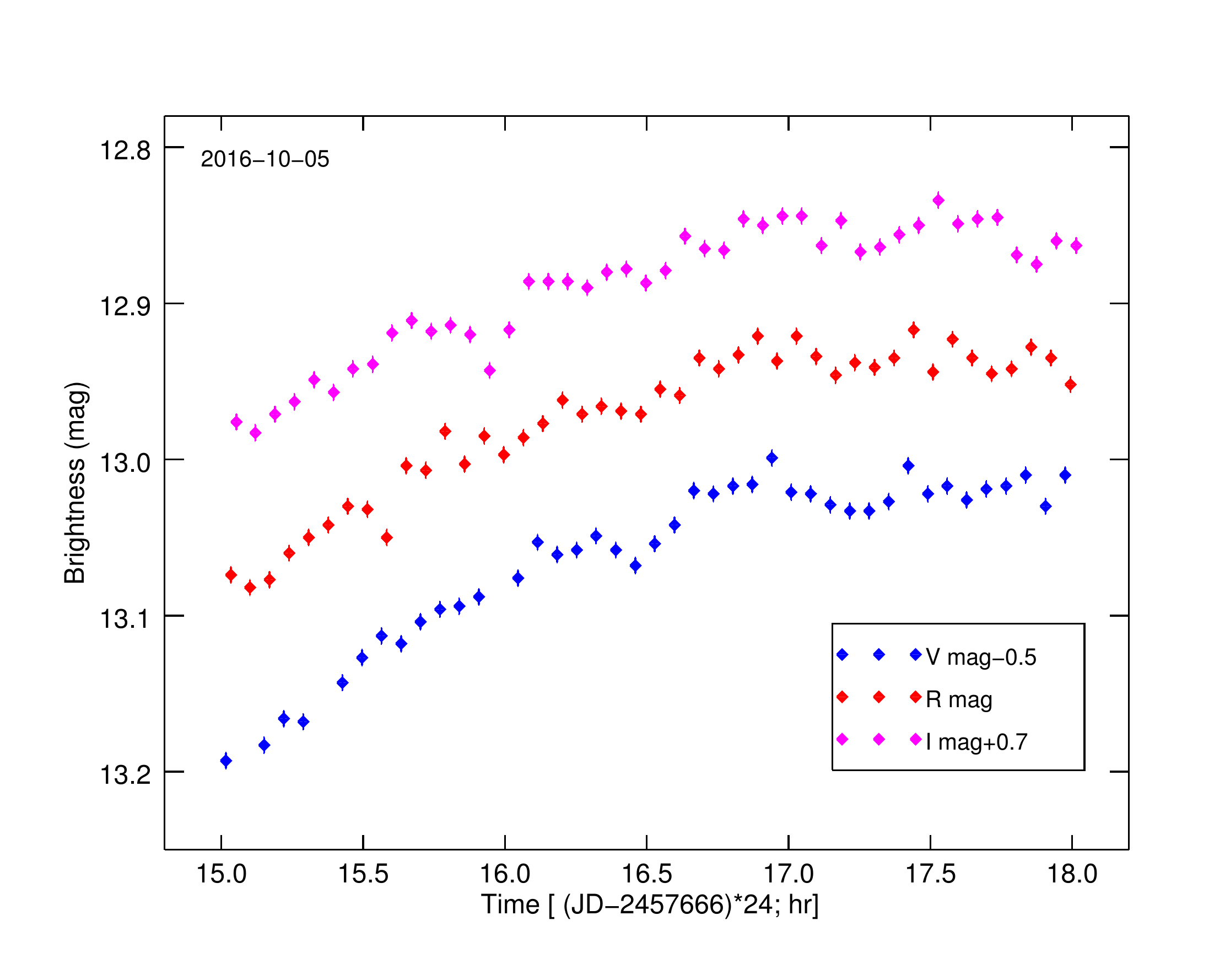}} \\
\end{tabular}
\end{center}
\caption{Multiband optical intranight observations of the blazar BL Lac. The observation dates and the filters  are specified on the plots.}
\label{LCs}
\end{figure}

\section{Analysis and Results}
To study the multiband optical variability properties of a sample of radio-loud sources, we have been frequently monitoring them for the past several years\footnote{For the list of the sources go to: http://faculty.fiu.edu/~webbj/micro\_web.html}. In general, the observations can be of the two types: single observation of the source for the purpose of long-term monitoring and microvariability observations i.e. continuously tracking the source throughout the night. Left panel of Figure \ref{fig:1} presents the multi-band light curves spanning $\sim5$ months showing the mean magnitude of the nightly photometric observations of BL Lac in the year 2016. The figure shows that  the source, displaying dramatic variability,  brightens by $\sim$1 mag ($\sim$2.5 times) in V and R band within a timescale of a month. The right panel of the figure shows the spectra in the optical bands (VRI) which can be approximated as power-law shapes with a mean spectral slope of 1.1. However, it can be seen that some of the spectra are markedly different from the power-law shapes, e. g. on 164th day, revealing significant spectral breaks.

In addition to the long-term flux points, we performed microvariability observed in 4 nights as mentioned earlier. The details of the multiband micorvariability analysis of the blazar BL Lac and the results following are discussed below.
\subsection{Multi-band micro-variability}
The intranight variability in the photometric bands  VRI exhibited by the source during the 4 nights are presented in Figure \ref{LCs}. The observation date, photometric filter, the mean magnitude and the observation duration are listed in the 1st, 2nd, 3rd and 4th column of Table 1, respectively. To quantify the observed variability, we estimated variability amplitude (VA) indicating peak-to-peak oscillation, and fractional variability (FV) representing mean variability.
 The amplitude of the peak-to-peak variations was estimated by using the relation given in \cite{Heidt1996},
\begin{equation}
{\rm VA} = \sqrt{(A_{max}-A_{min})^2-2\sigma ^{2}} \, , 
\end{equation}
where $A_{max}$, $A_{min}$, and $\sigma$ are the maximum, minimum, and the mean of the magnitude errors in the light curves, respectively. Similarly for a mean flux of $\left \langle F \right \rangle$ with $ S^{2}$ variance and $\left \langle \sigma _{err}^{2} \right \rangle$ mean squared uncertainties, the fractional variance, as in \citep{vau03}, is given as 
\begin{equation}
F_{var}=\sqrt{\frac{S^{2}-\left \langle \sigma _{err}^{2} \right \rangle}{\left \langle F \right \rangle^{2}}} .
\end{equation}
The error in the fractional variability can be expressed as
\begin{equation}
\centering
\sigma_{F_{var}}=\sqrt{ F_{var}^{2}+\sqrt{ \frac{2}{N}\frac{\left \langle \sigma _{err}^{2} \right \rangle^{2}}{\left \langle F \right \rangle^{4}}+  \frac{4}{N}\frac{\left \langle \sigma _{err}^{2} \right \rangle }{\left \langle F \right \rangle^{2}} F _{var}^{2}}} - F_{var}
\end{equation}
 \citep[see][]{Aleksic2015}.
The multiband VA and FV for all the observations are  listed in the 5th and 6th column of the Table 1, respectively.  These quantities show that the source displays multiband fast variability by changing the flux by as large as $\sim0.2$ R mag (or $\sim20\%$) within a timescale of $\sim3$ hrs. In addition, we note that for the given night, both VA and $F_{\rm var}$ are consistently larger for the higher energy optical band implying larger variability for the higher frequency emission.

\begin{table}[t!]
\caption{Summary of the microvariability program in the year 2016.}
\centering
\begin{tabular}{cccccc}
\toprule
Date	&	Band	&	Mean Mag.	&	Duration (hr)	&	VA (mag)	&	F$_{var}$ (\%)	\\
\hline
2016-08-07	&	R	&13.48	$\pm$	0.002		&4.3		&0.08		&1.47$\pm$0.03		\\
2016-08-12	&	R	&13.359	$\pm$	0.002		&3.8		&0.11		&3.37$\pm$0.05		\\
2016-08-12	&	V	&13.986	$\pm$	0.003	&3.8		&0.13		&3.70$\pm$0.05		\\
2016-09-30	&	V	&14.099	$\pm$	0.018		&6.2		&0.08		&1.61$\pm$0.03		\\
2016-09-30	&	I	&12.670	$\pm$	0.003		&6.2		&0.07		&1.35$\pm$0.03		\\
2016-10-05	&	V	&13.5640$\pm$	0.003		&2.9		&	0.19	&	4.75$\pm$0.07	\\	
2016-10-05	&	R	&12.970	$\pm$	0.002		&2.9		&0.17		&4.30$\pm$0.07		\\
2016-10-05	&	I	&12.193	$\pm$	0.002		&2.9		&0.14		&3.79$\pm$0.07		\\

\bottomrule
\end{tabular}
\end{table}

  \begin{figure}[t!]
\centering
\begin{center}
\begin{tabular}{c@{}c@{}c}
\resizebox{0.33\textwidth}{!}{\includegraphics{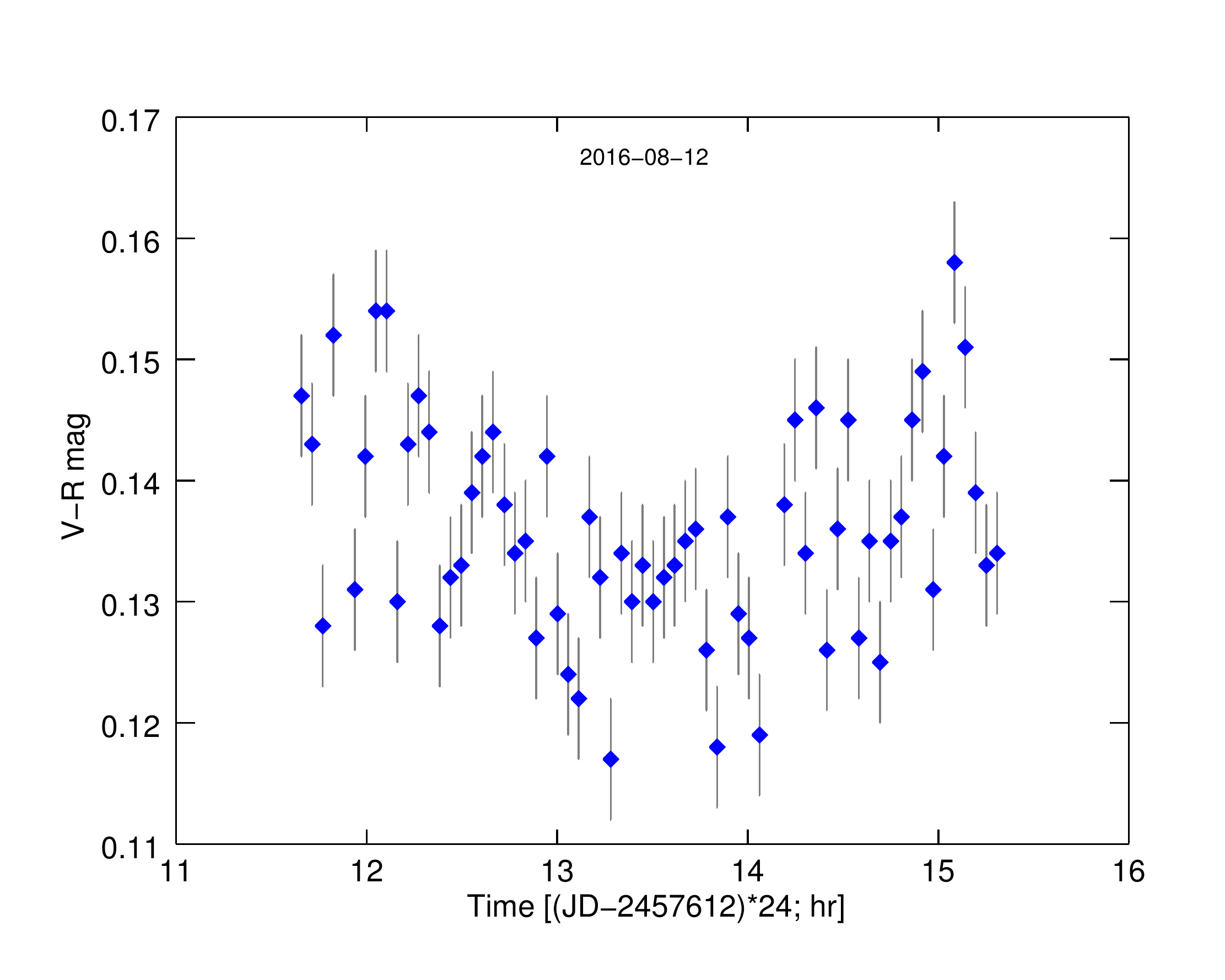}} &
\resizebox{0.33\textwidth}{!}{\includegraphics{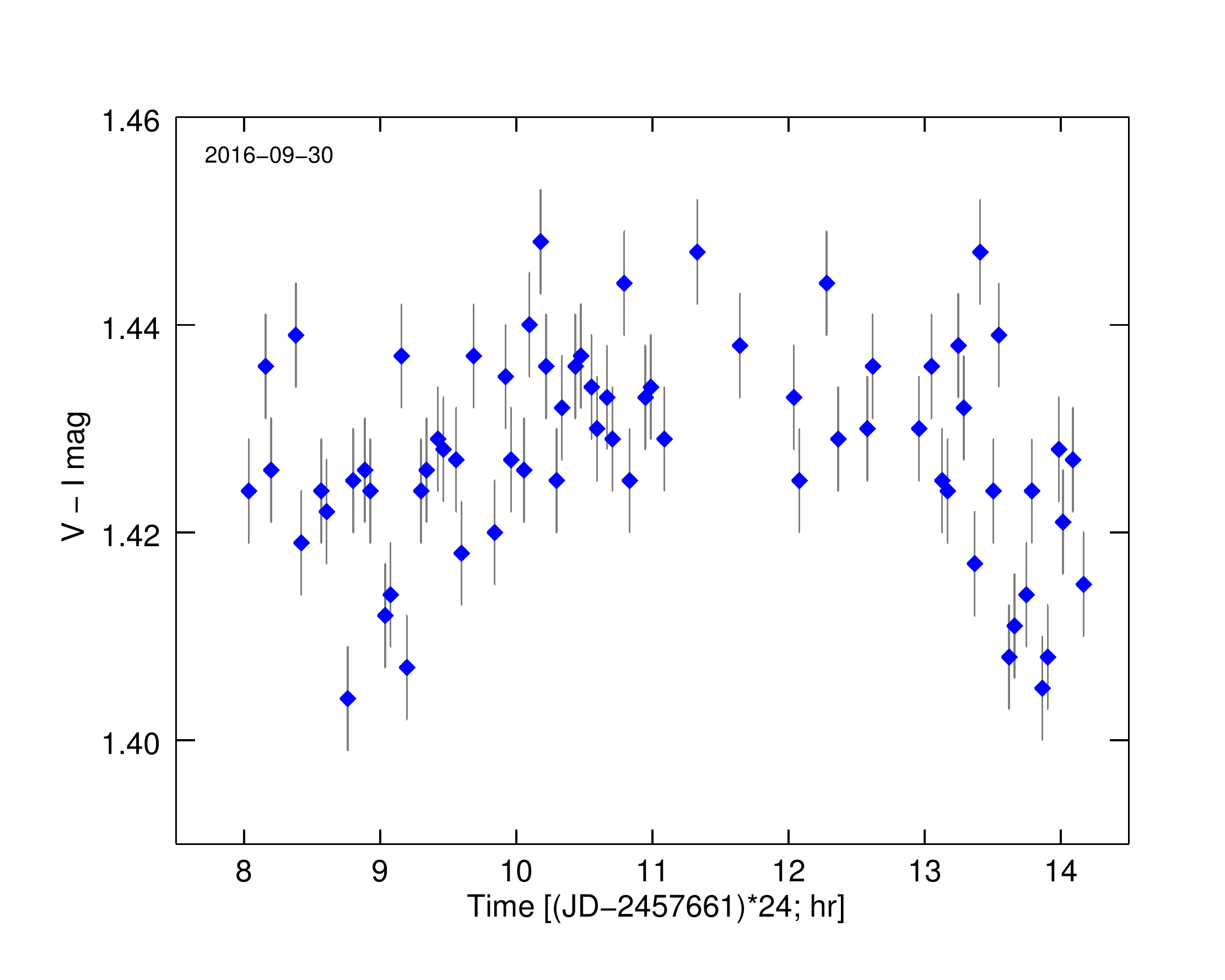}} &
\resizebox{0.33\textwidth}{!}{\includegraphics{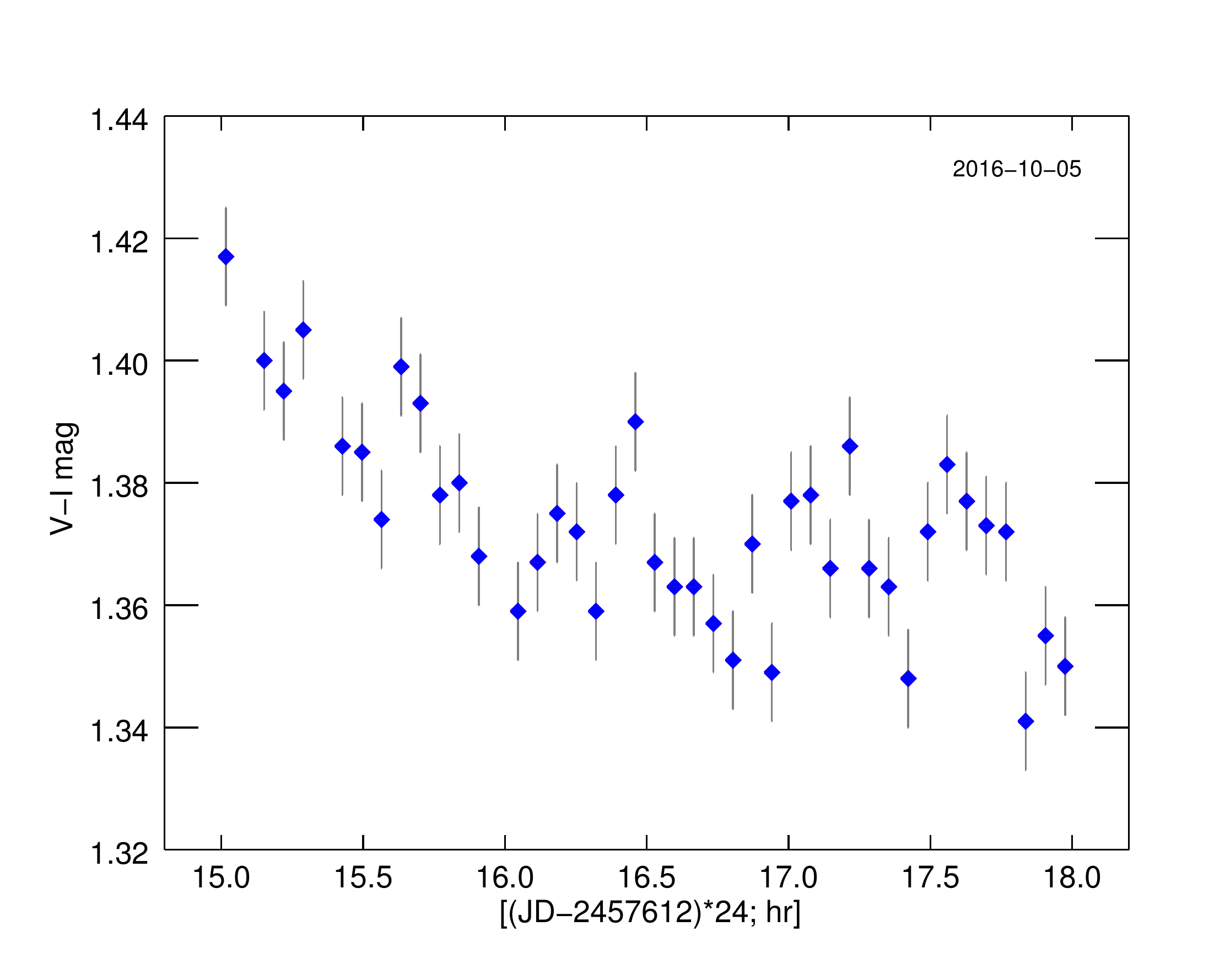}} \\
\end{tabular}
\end{center}
\caption{Intranight optical color (V-I or V-R) variability of the blazar BL Lac}
\label{color}
\end{figure}

\begin{figure}[b!]
\centering
\begin{center}
\begin{tabular}{c@{}c@{}c}
\resizebox{0.33\textwidth}{!}{\includegraphics{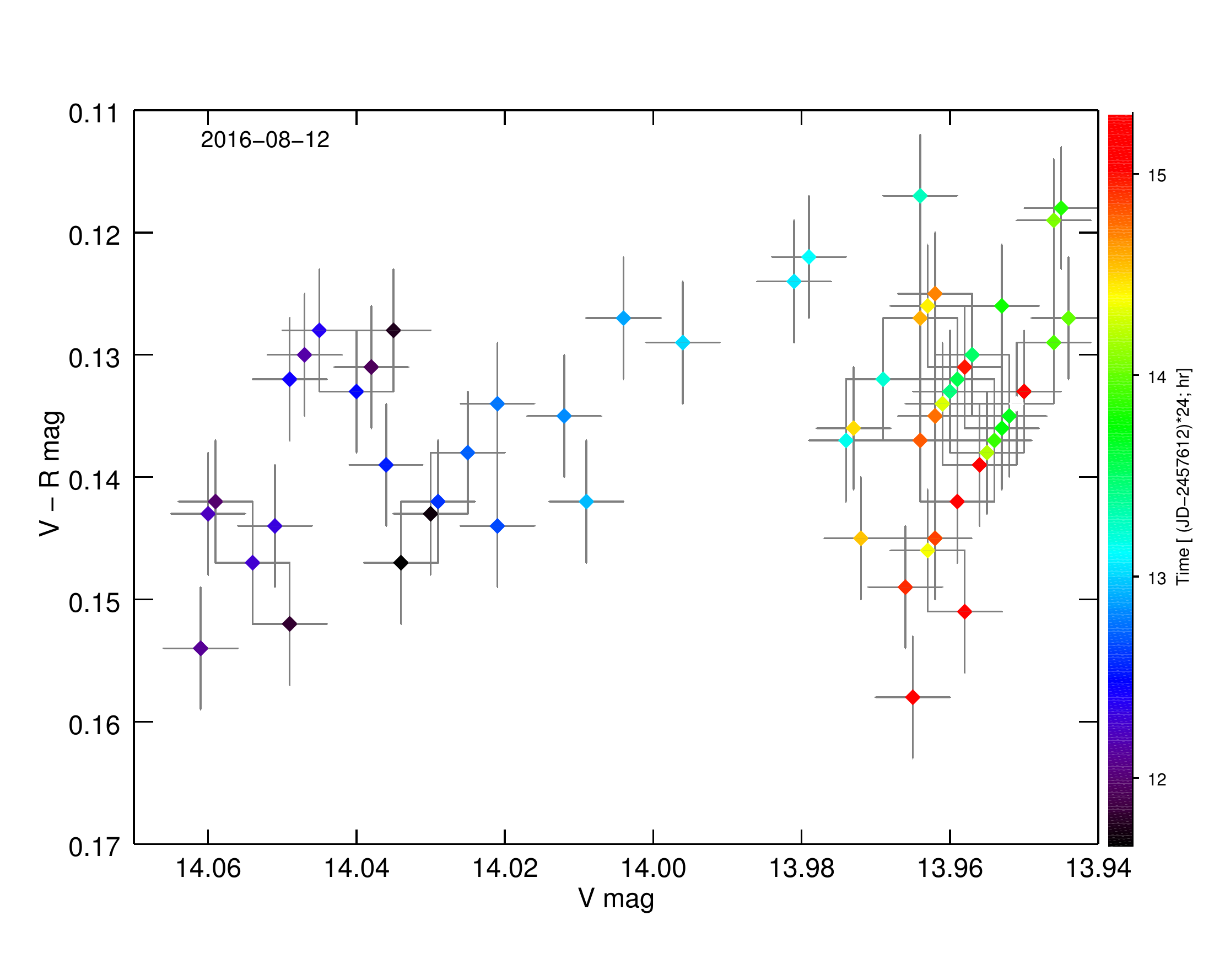}} &
\resizebox{0.33\textwidth}{!}{\includegraphics{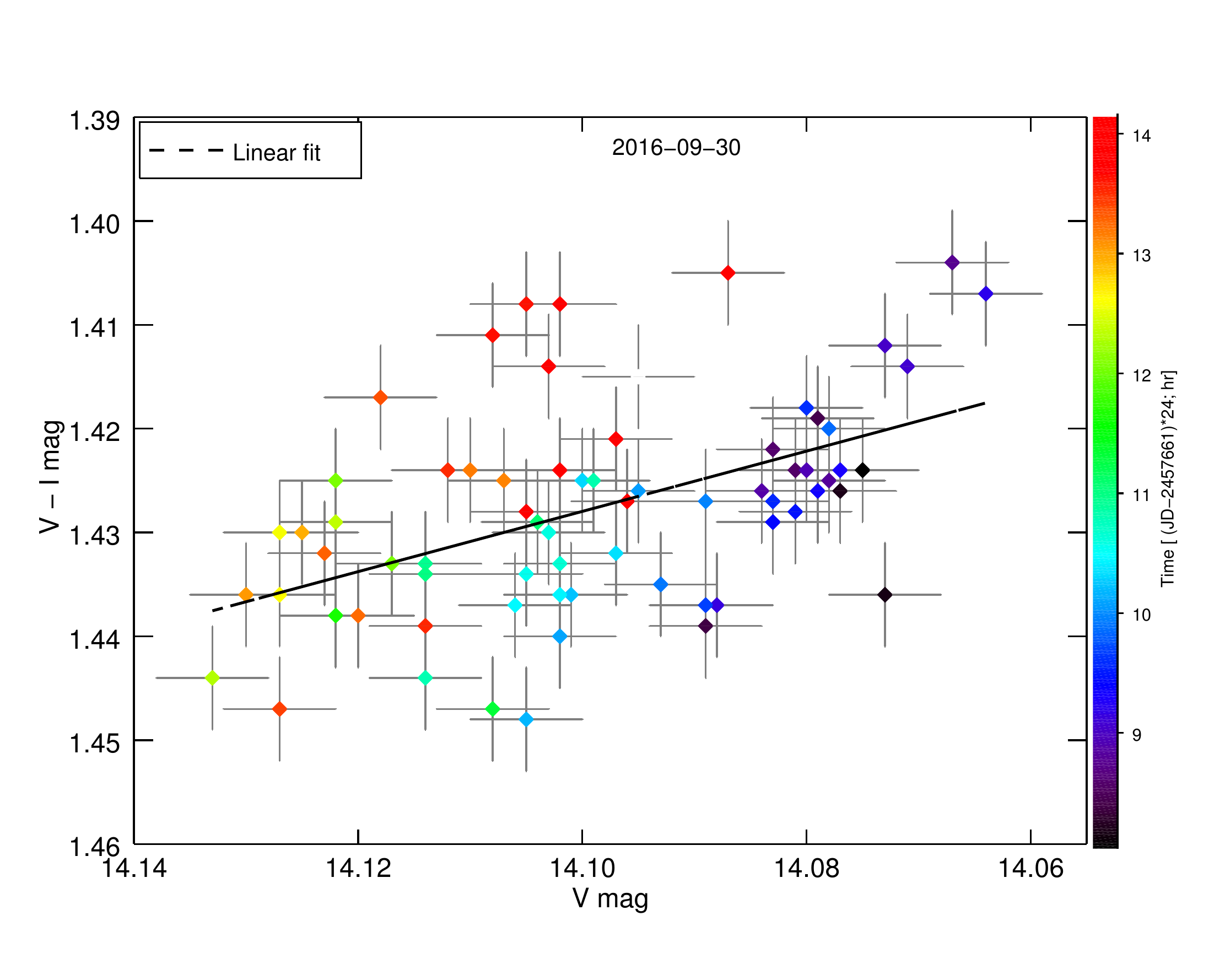}} &
\resizebox{0.33\textwidth}{!}{\includegraphics{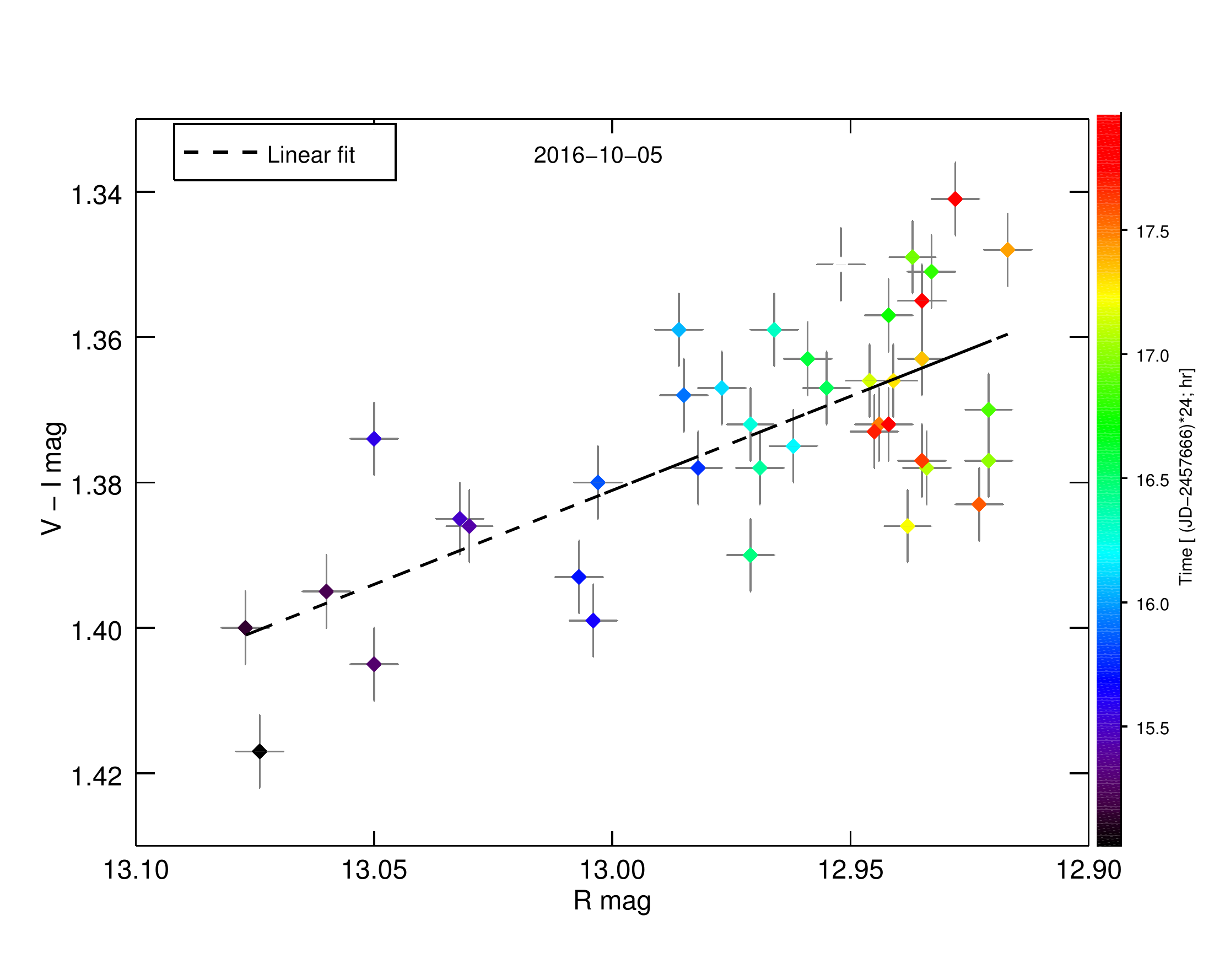}} \\
\end{tabular}
\end{center}
\caption{ Intranight optical  color magnitude relation the blazar BL Lac. The symbols are color coded to represent time.}
\label{color-mag}
\end{figure}

\subsubsection{Color variability: Bluer-when-brighter}
The relative source flux changes in different bands give rise to the intranight color evolution. Here color refers to the quasi-simultaneous (simultaneous within 1 minute) difference in the magnitudes in V and I (or R)  filters. Figure \ref{color} shows that, along with flux variability, the source also exhibits considerable color  variability ($\sim0.05-0.20$ mag) within a few hours.  To study the relation between the flux in magnitudes and color variations,  colors are plotted against the corresponding V ( R for Night 3)  band magnitudes. The relation can be termed as  bluer-when-bright trend which appears less, moderately and strongly pronounced during Night 2, Night 3 and Night 4, respectively. Similar observations were made by \citep{Zhang2013} in their study of the source on intranight timescales .
\subsection{Multi-band cross-correlation}
An investigation into the relation among emission in different energy bands can offer an important insight regarding the structure of the blazar emission regions, the dominant radiative processes involved and distribution of the emitting particles.  The discrete correlation function (DCF) described in  \cite{EK88}, is one of the most extensively used methods to investigate the cross-correlation between two time series with uneven spacing  \citep[see][]{bhatta016}. The unbinned DCF can be estimated as,
\begin{equation}
UDCF_{ij}=\frac{\left ( x_{i}-\bar{x} \right )\left ( y_{j}-\bar{y} \right )}{\sqrt{\left ( \sigma _{x}^{2} -e_{x}^{2}\right )\left ( \sigma _{y}^{2} -e_{y}^{2}\right )}}
\label{UDCF}
\end{equation}
where $\bar{x}$ and $\bar{y}$, $\sigma^{2}$, $e^{2}$ correspond to the mean, variance and the uncertainties in the magnitudes of the two light curves, respectively. These discrete pairs can be binned of bin width comparable to the sampling widths of the light curves. Then the average DCF including the M pairs within the given bin width is written as, 
\begin{equation}
DCF(\tau )=\frac{1}{M}UDCF_{ij}
\label{DCF}
\end{equation}

\begin{figure}[t!]
\centering
\begin{center}
\begin{tabular}{c@{}c@{}c}
\resizebox{0.33\textwidth}{!}{\includegraphics{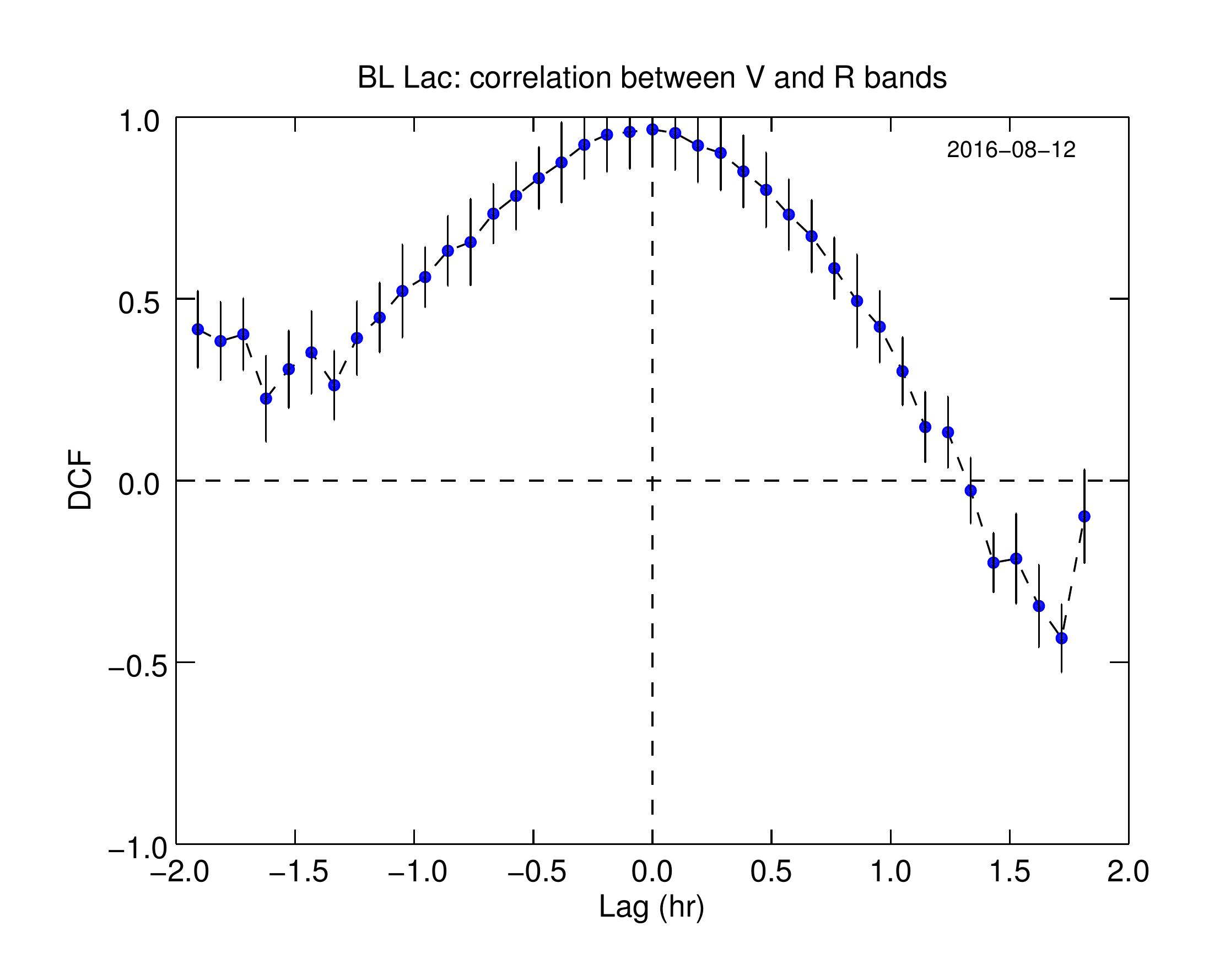}} &
\resizebox{0.33\textwidth}{!}{\includegraphics{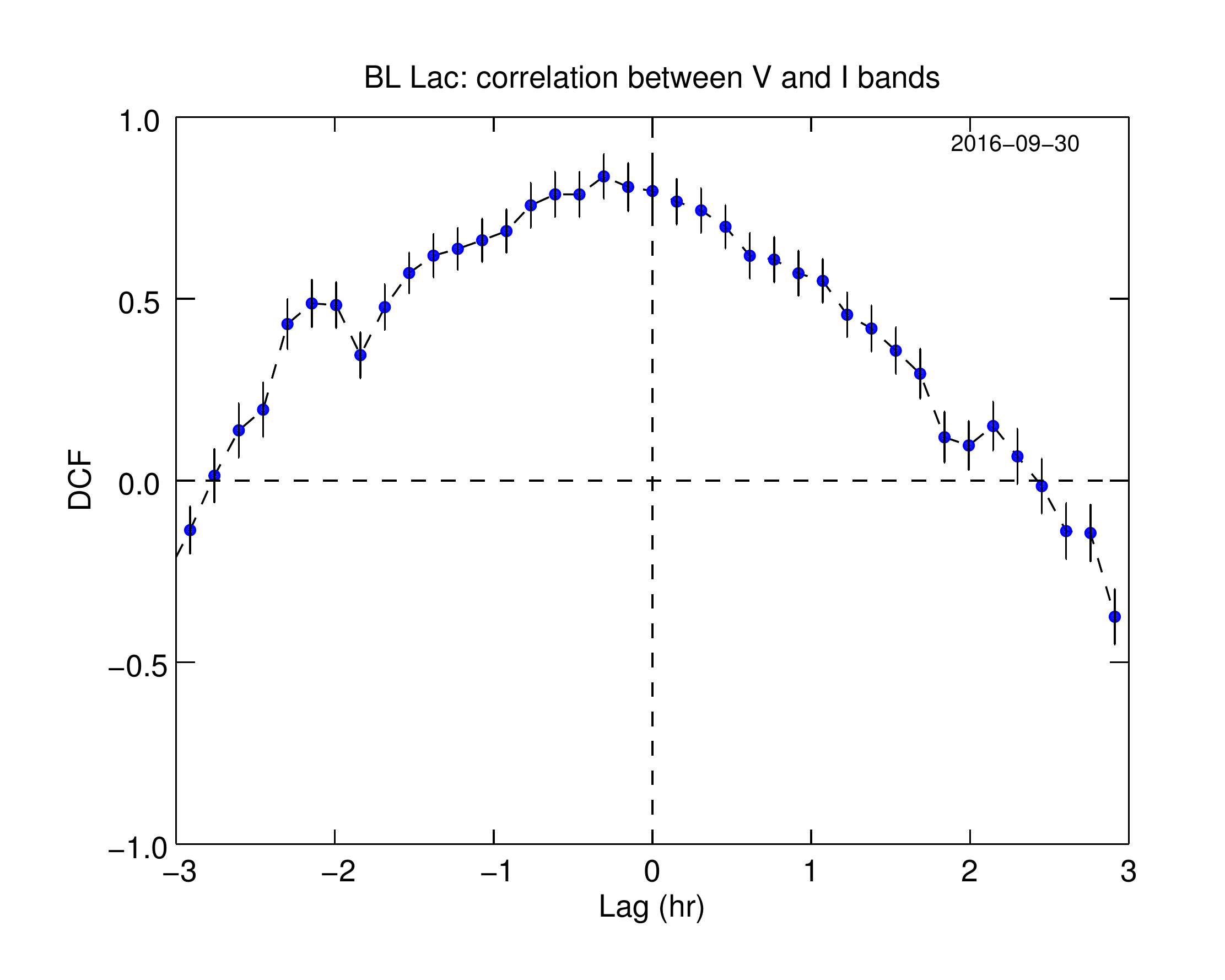}}&
\resizebox{0.33\textwidth}{!}{\includegraphics{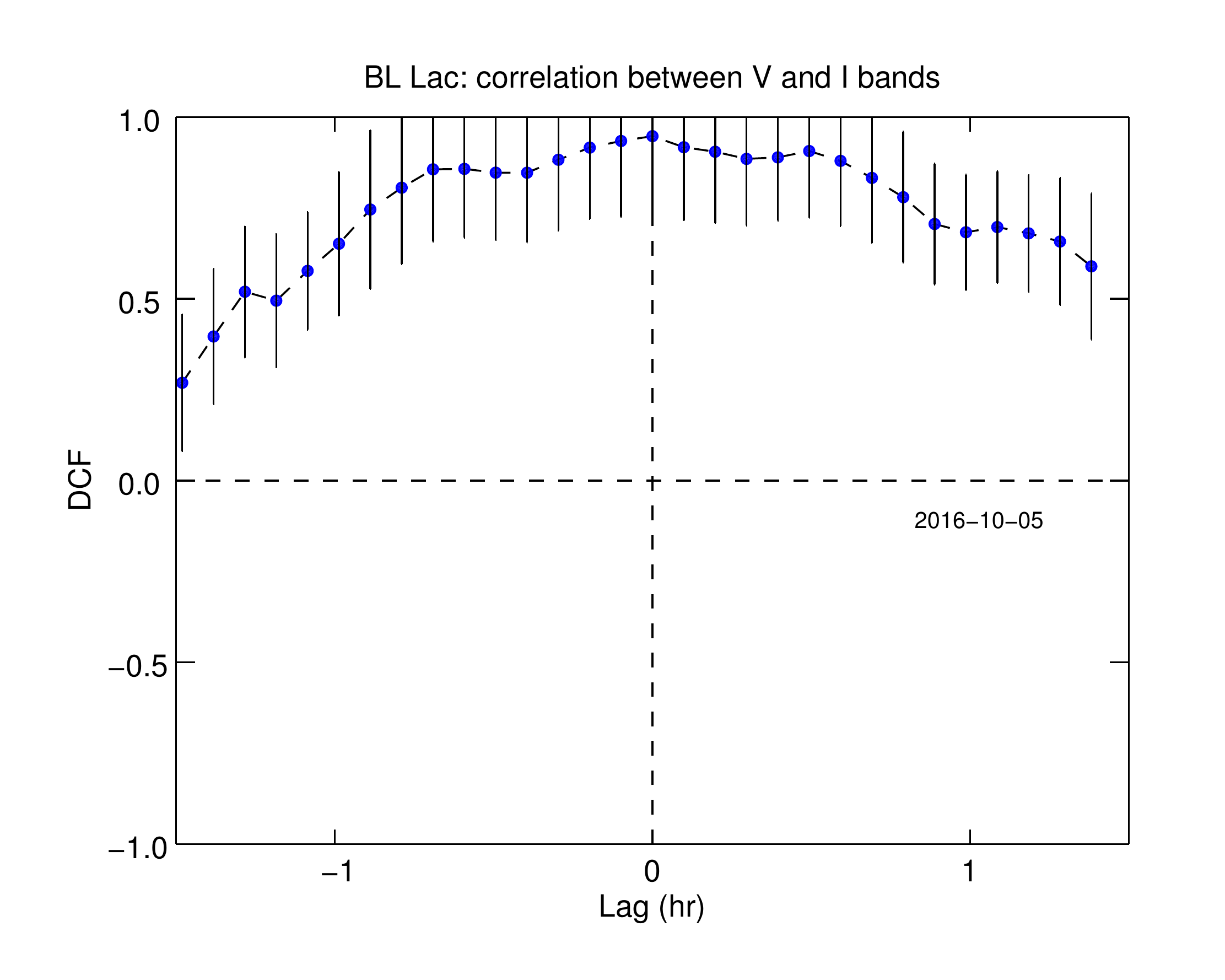}} \\
\end{tabular}
\end{center}
\caption{Discrete cross-correlation function between the muliband emission of the blazar BL Lac on intranight timescales.}
\label{DCF1}
\end{figure}

 We estimated the DCF between  V and R band observations during Night 2, and between V and I band observation during  Night 3 and Night 4. For  Night 2 and Night 4 the emission in the two bands were found to be highly correlated at the zero lag. But for Night 3 a soft lag (I band emission lagging behind V band emission) of $\sim 20$ minutes was witnessed as shown in Figure \ref{DCF1}.

\subsubsection{Search for Characteristic timescales}
\begin{figure}[ht!]
\centering
\begin{center}
\begin{tabular}{c@{}c}
\resizebox{0.43\textwidth}{!}{\includegraphics{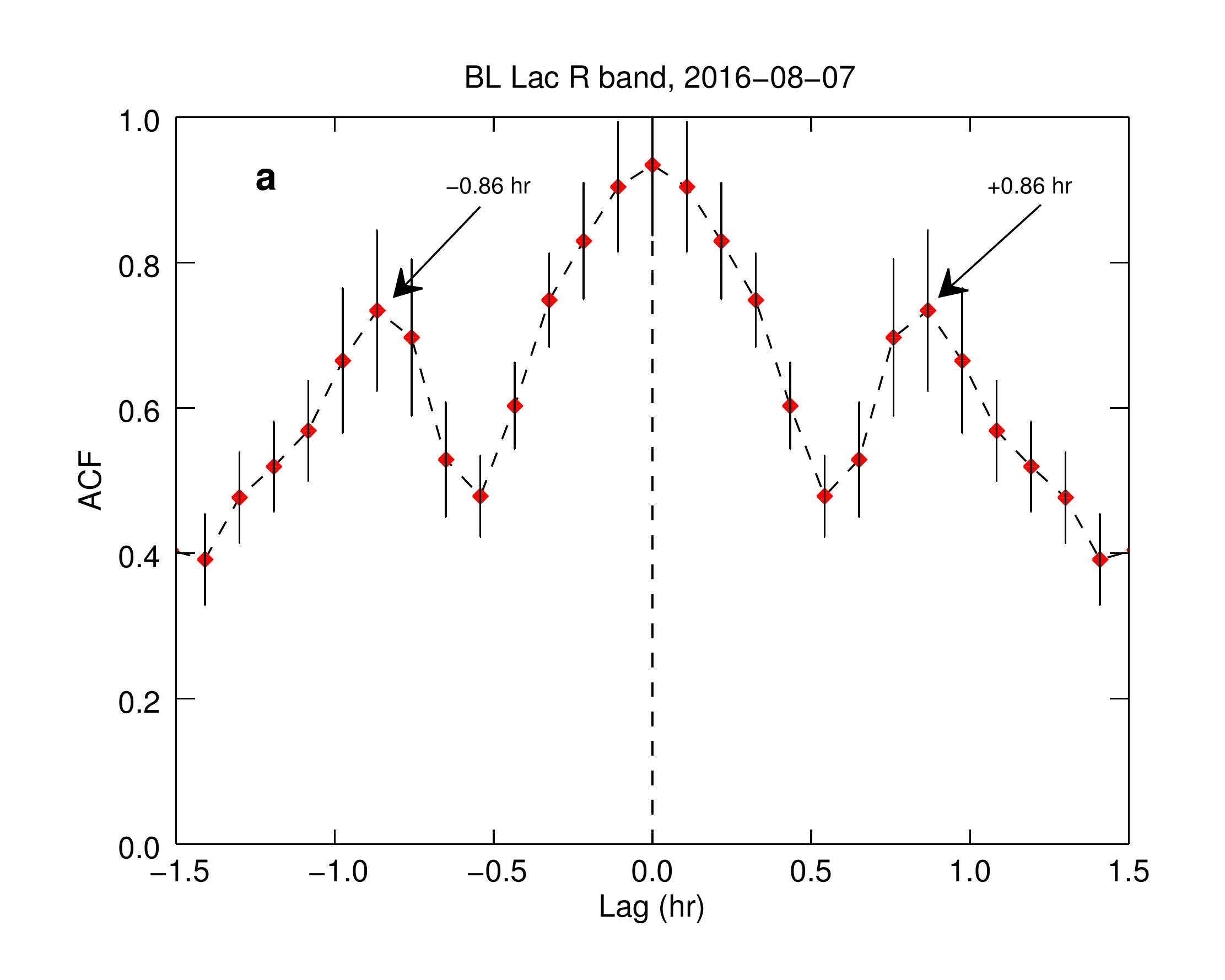}} &
\resizebox{0.43\textwidth}{!}{\includegraphics{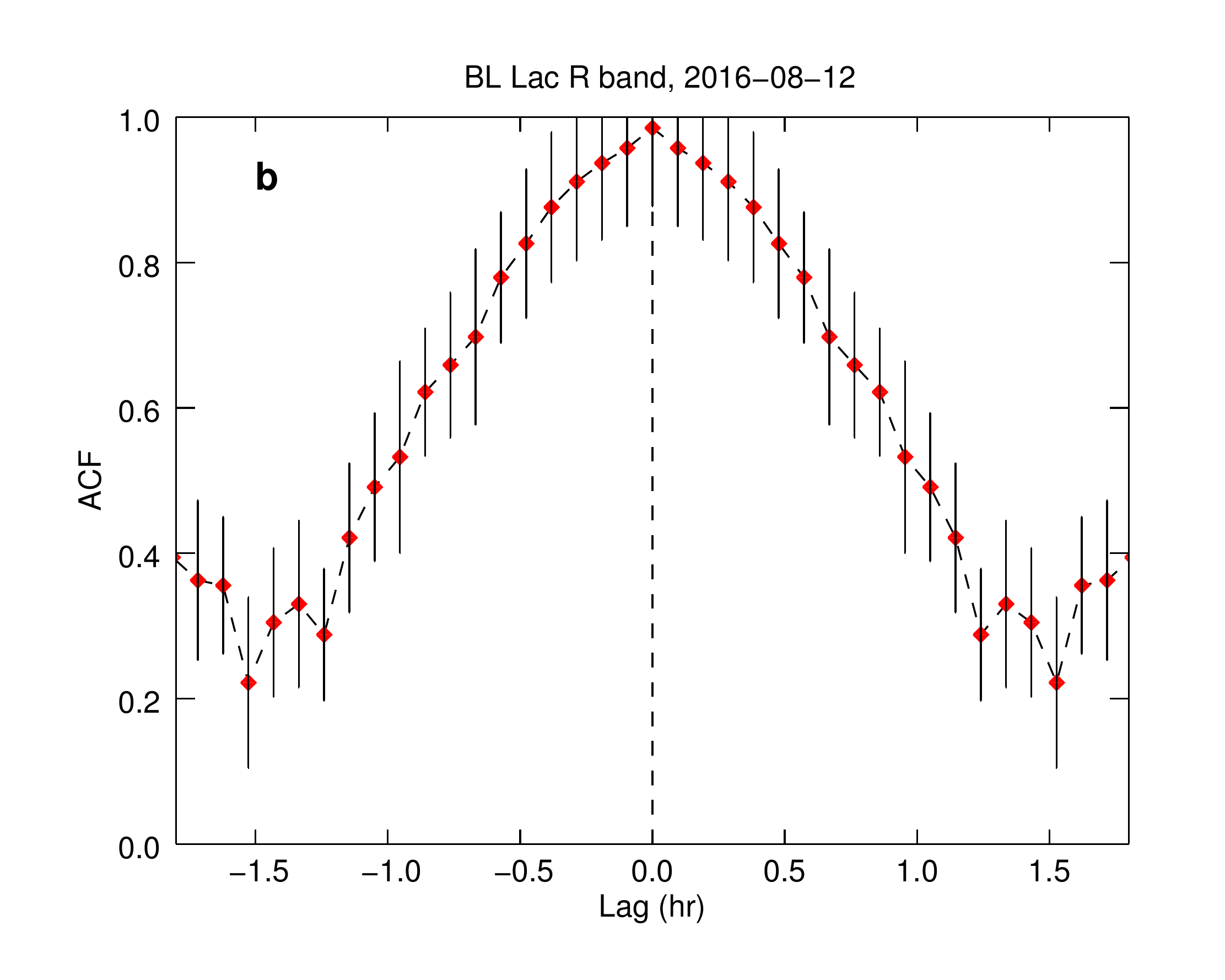}} \\
\resizebox{0.43\textwidth}{!}{\includegraphics{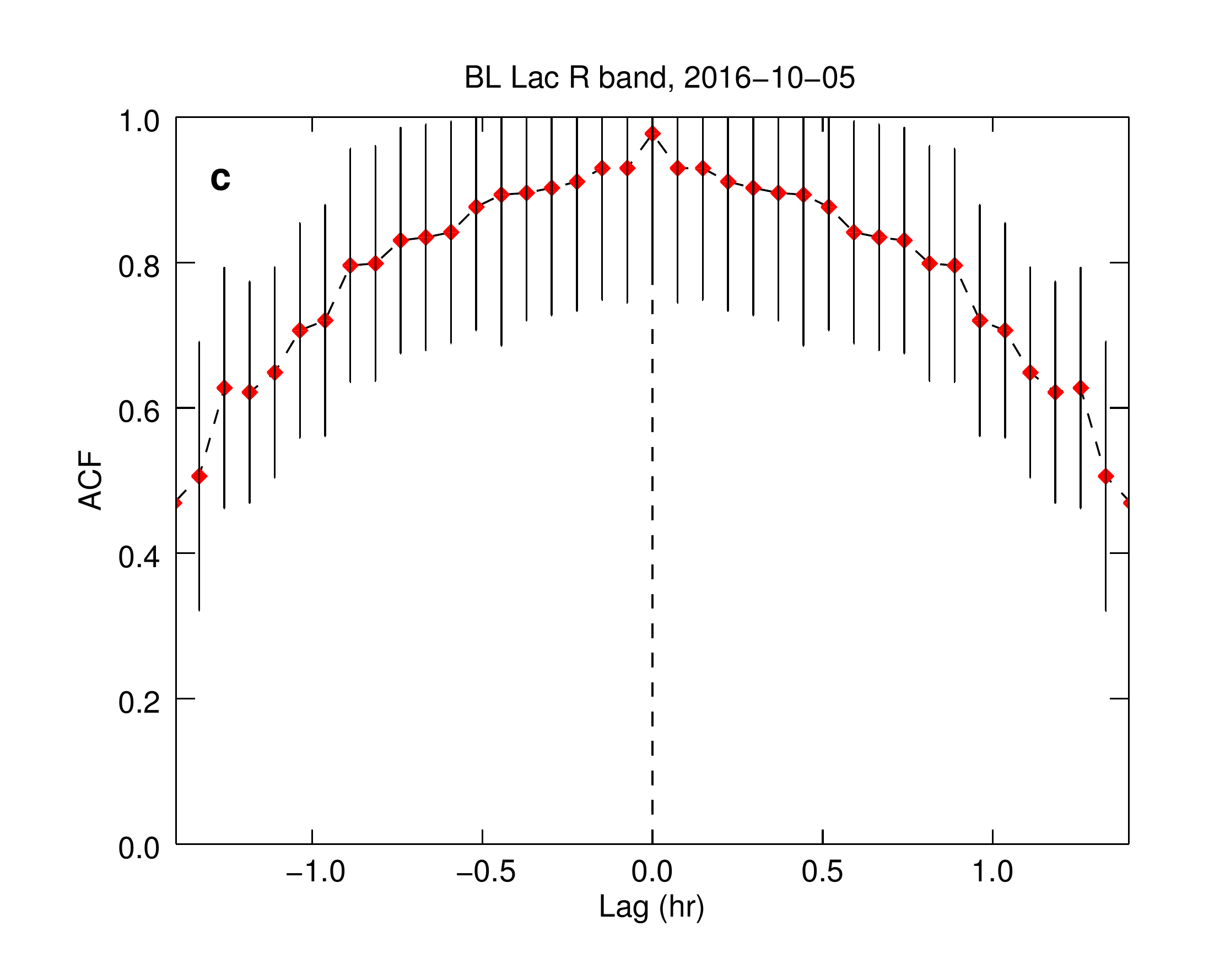}} &
\resizebox{0.43\textwidth}{!}{\includegraphics{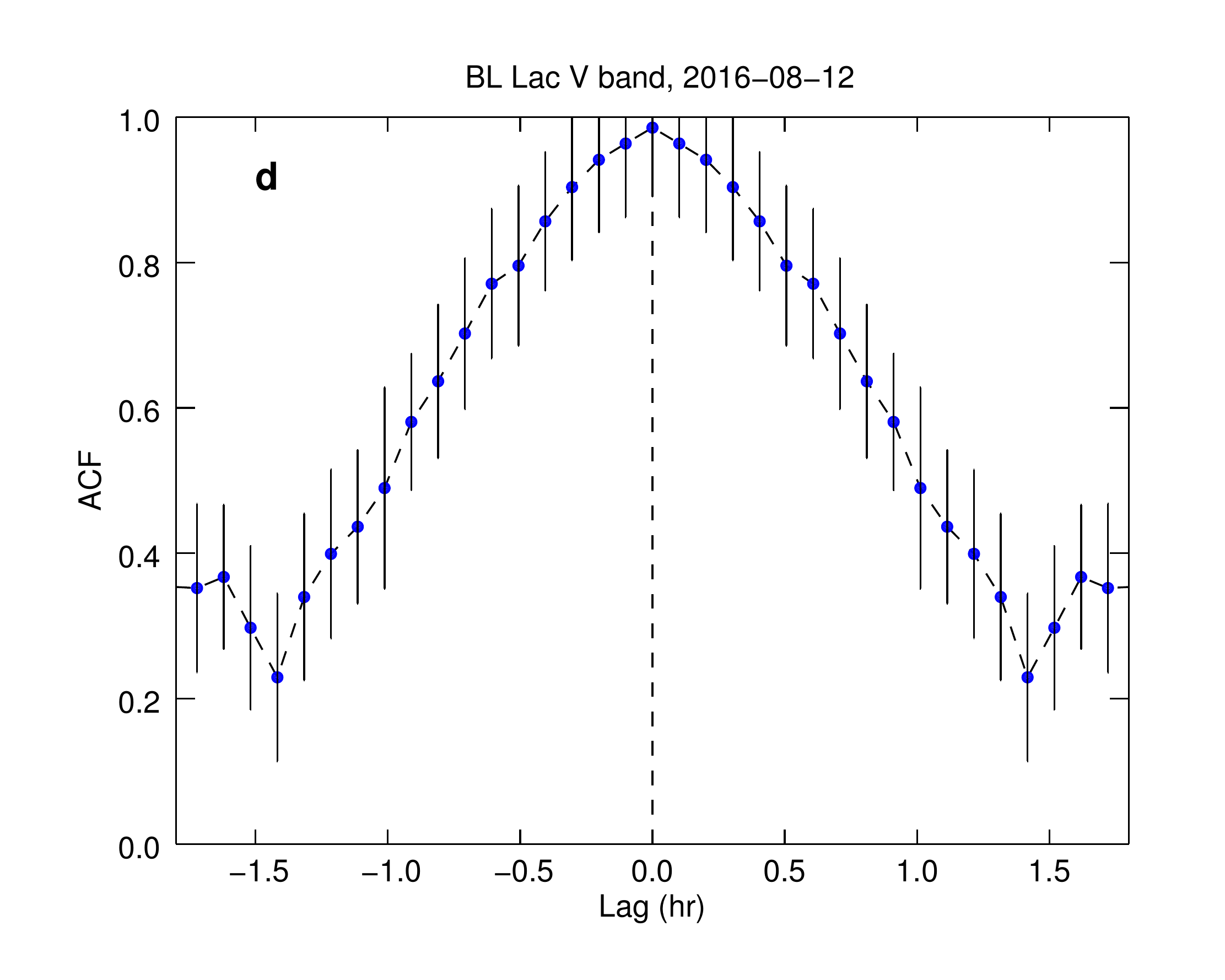}} \\
\resizebox{0.44\textwidth}{!}{\includegraphics{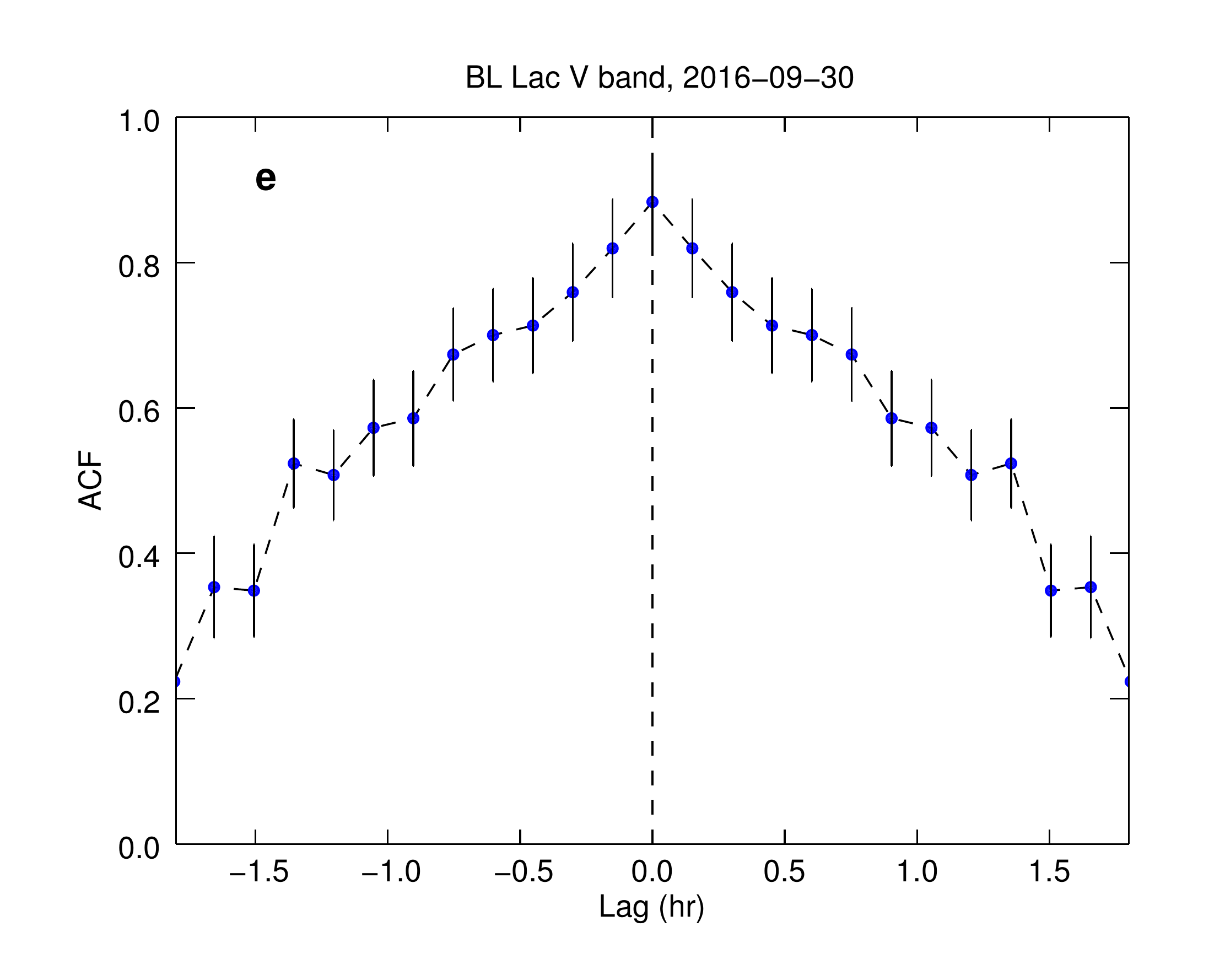}} &
\resizebox{0.44\textwidth}{!}{\includegraphics{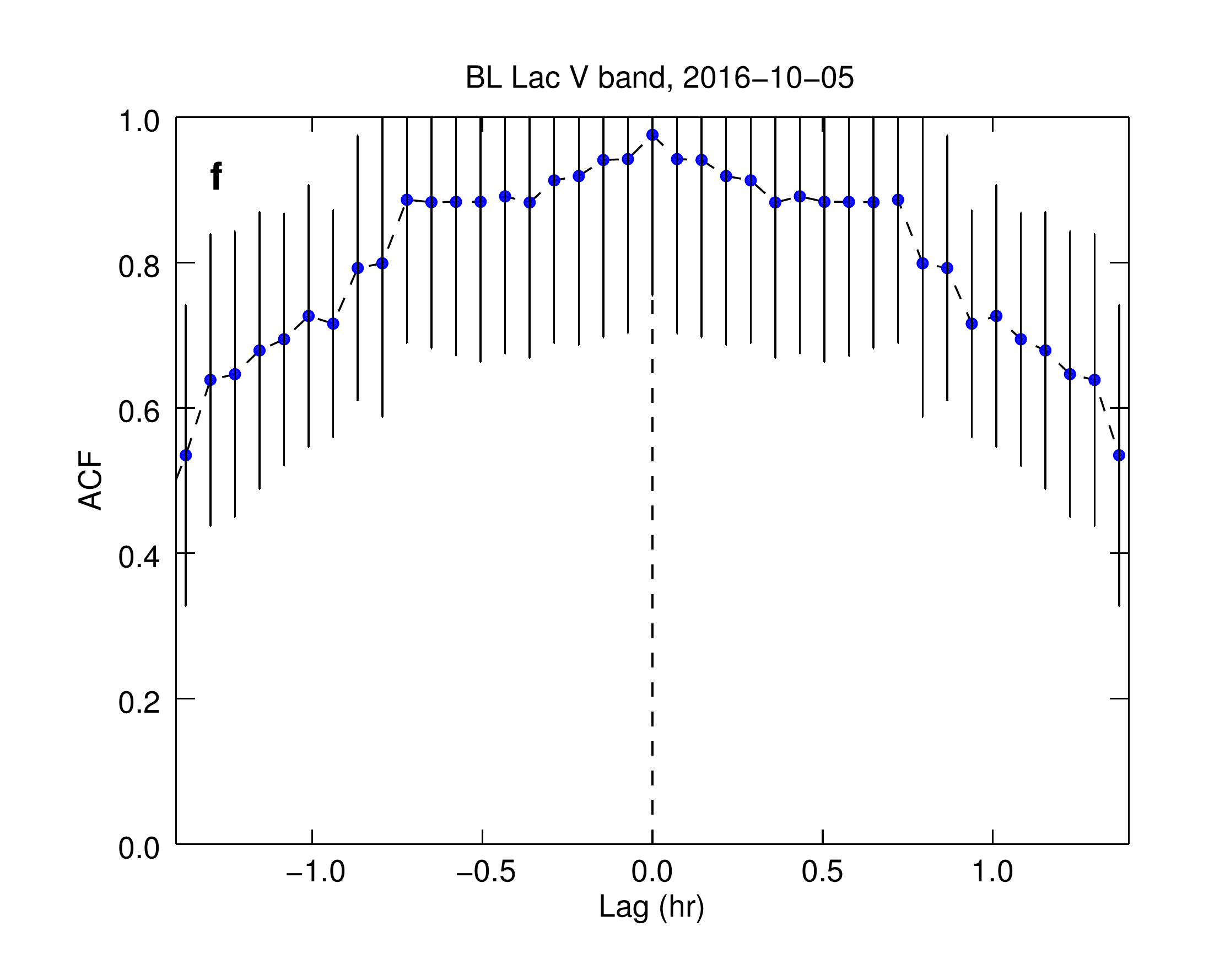}} \\
\resizebox{0.44\textwidth}{!}{\includegraphics{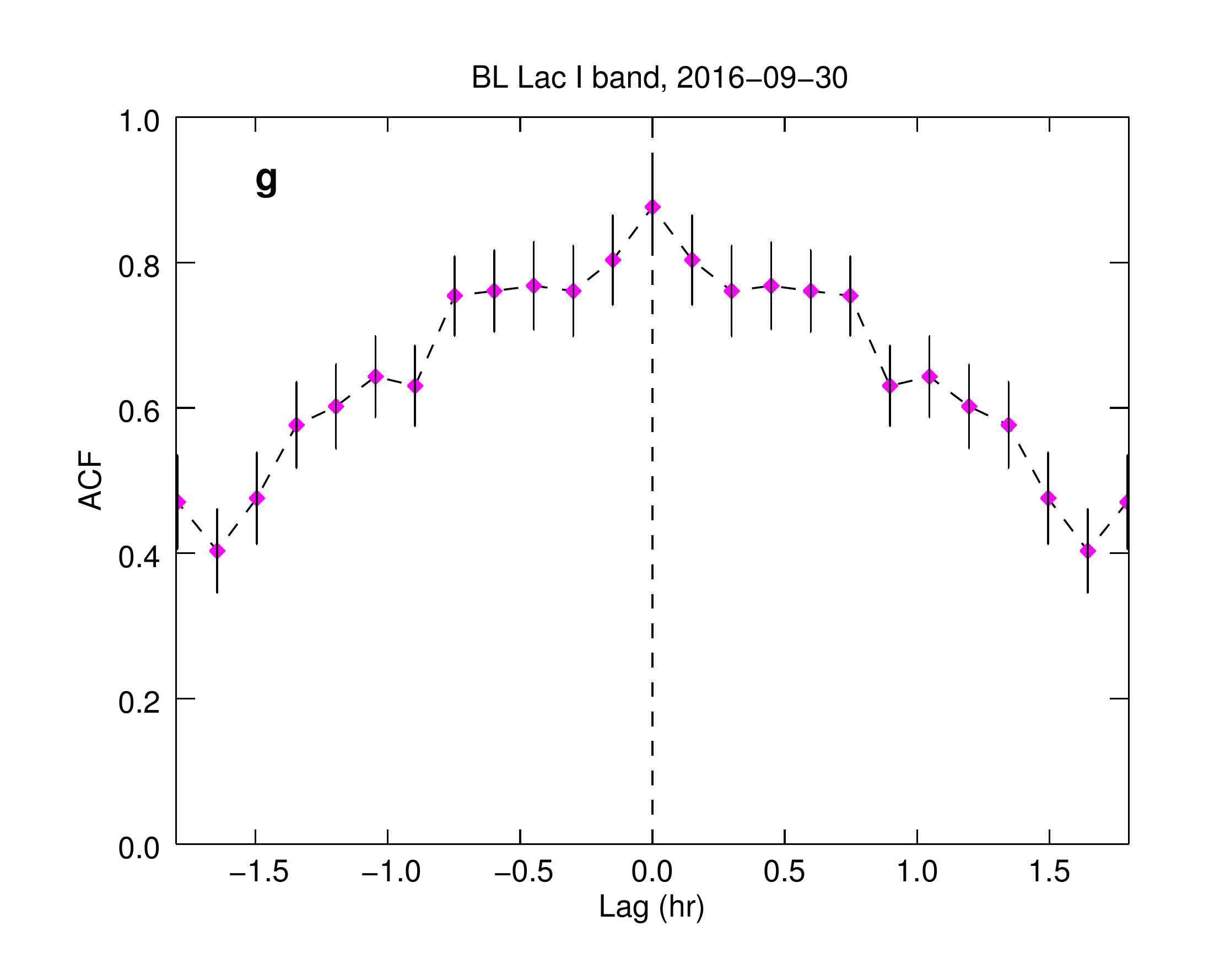}} &
\resizebox{0.44\textwidth}{!}{\includegraphics{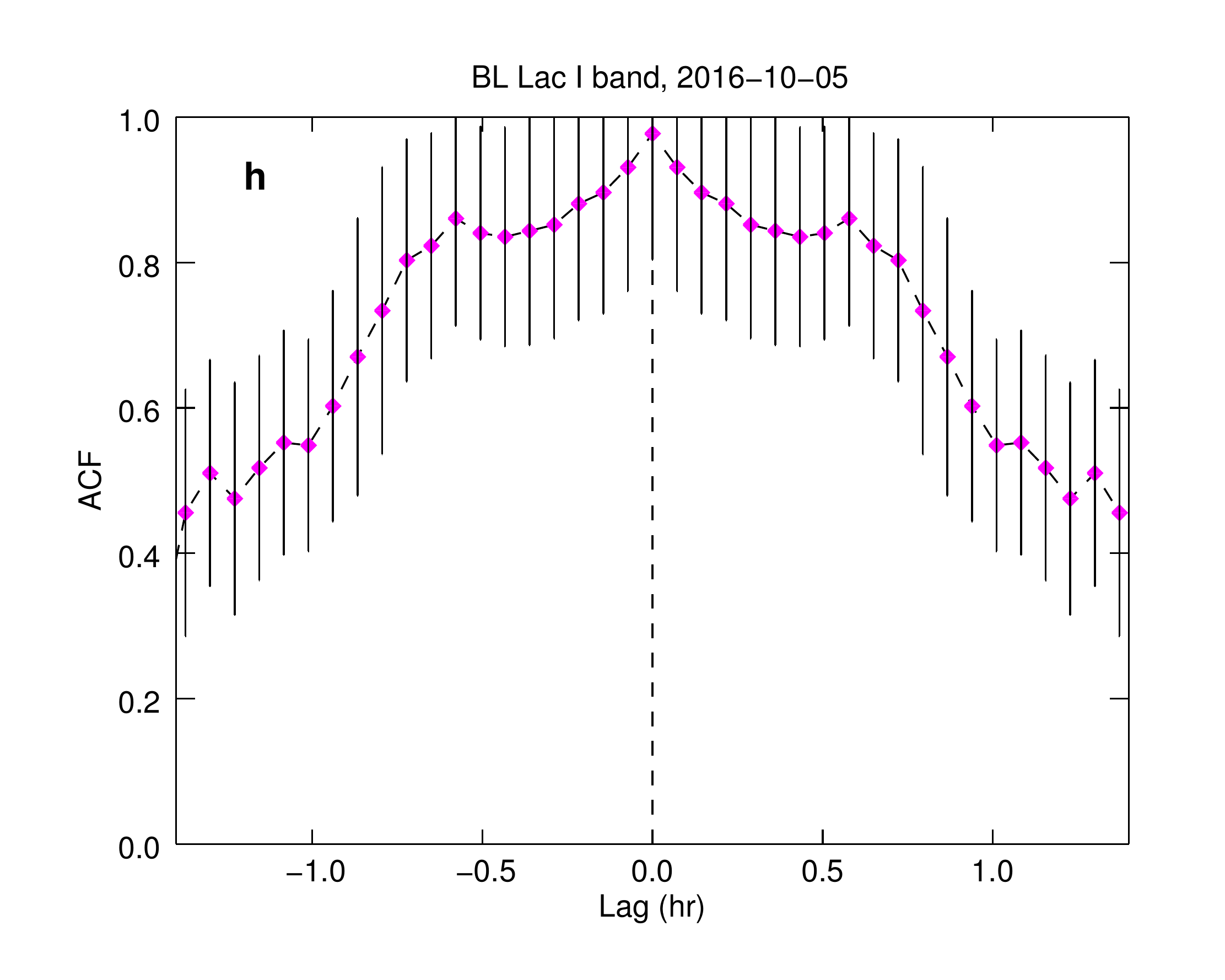}} \\
\end{tabular}
\end{center}
\caption{Intranight discrete auto-correlation function of the optical light curves of the blazar BL Lac}
\label{ACF}
\end{figure}

To search for the possible characteristic timescales present in the multiband light curves of the blazar BL Lac, we employed two widely used methods of time series: discrete auto-correlation function (ACF) and structure function (SF). Similar studies of blazars involving Fermi/LAT light curves were presented by \cite{Abdo2010}; and  \cite{Villata2004} 
used ACF and SF to find the possible periodicities in the longterm radio light curve of the source. An expression for the discrete auto-correlation function (ACF) can be obtained by setting x=y in the expressions given by Equations \ref{UDCF} and \ref{DCF}. The ACFs for all 8 intranight light curves are presented in the 8 panels of Figure \ref{ACF}. The figure shows presence of characteristic timescales in the R band light curve on Night 1 (Panel a) as represented by the well-resolved secondary peaks of at ($\sim |0.86|$ hr).

\begin{figure}[ht!]
\centering
\begin{center}
\begin{tabular}{c@{}c}
\resizebox{0.43\textwidth}{!}{\includegraphics{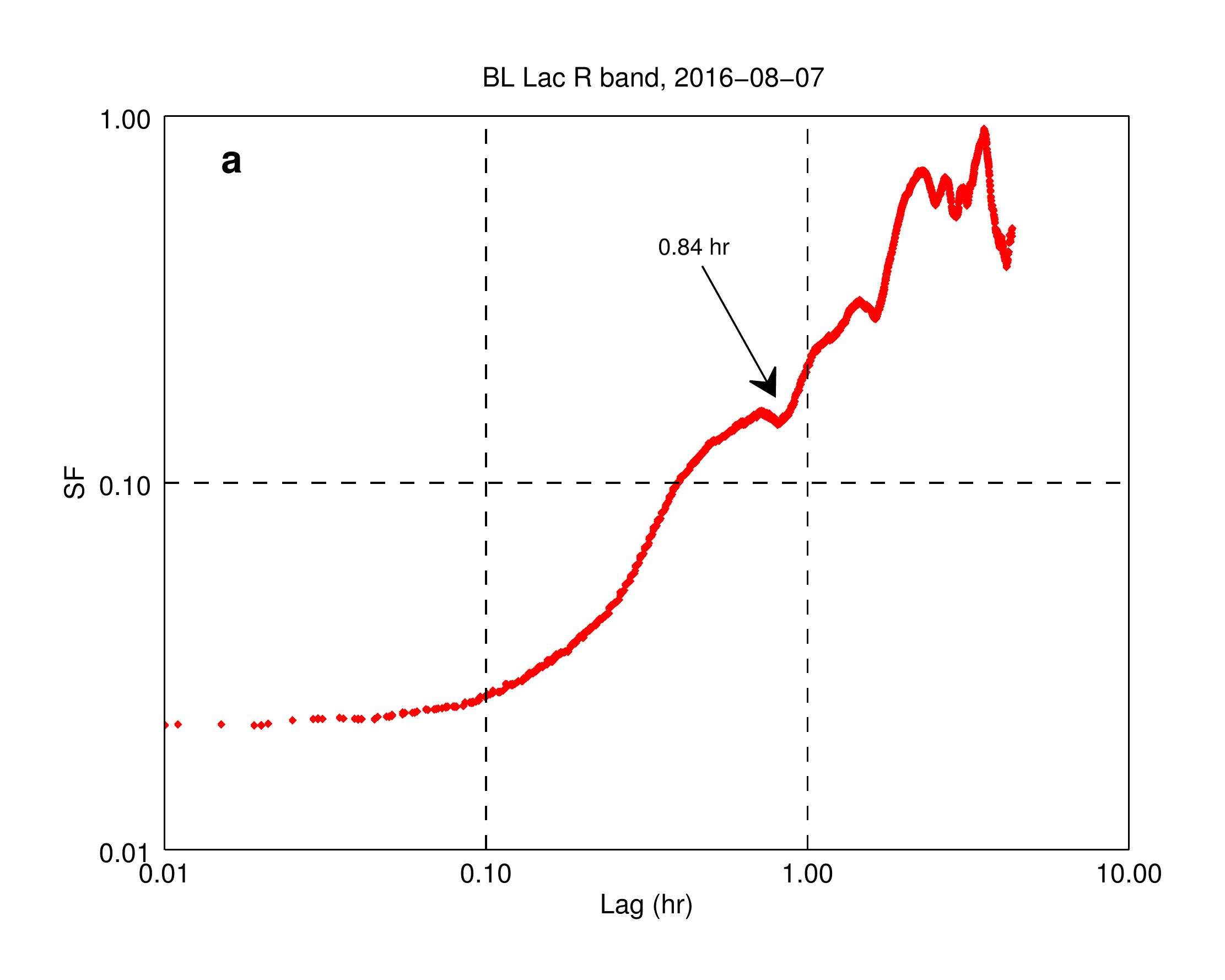}} &
\resizebox{0.43\textwidth}{!}{\includegraphics{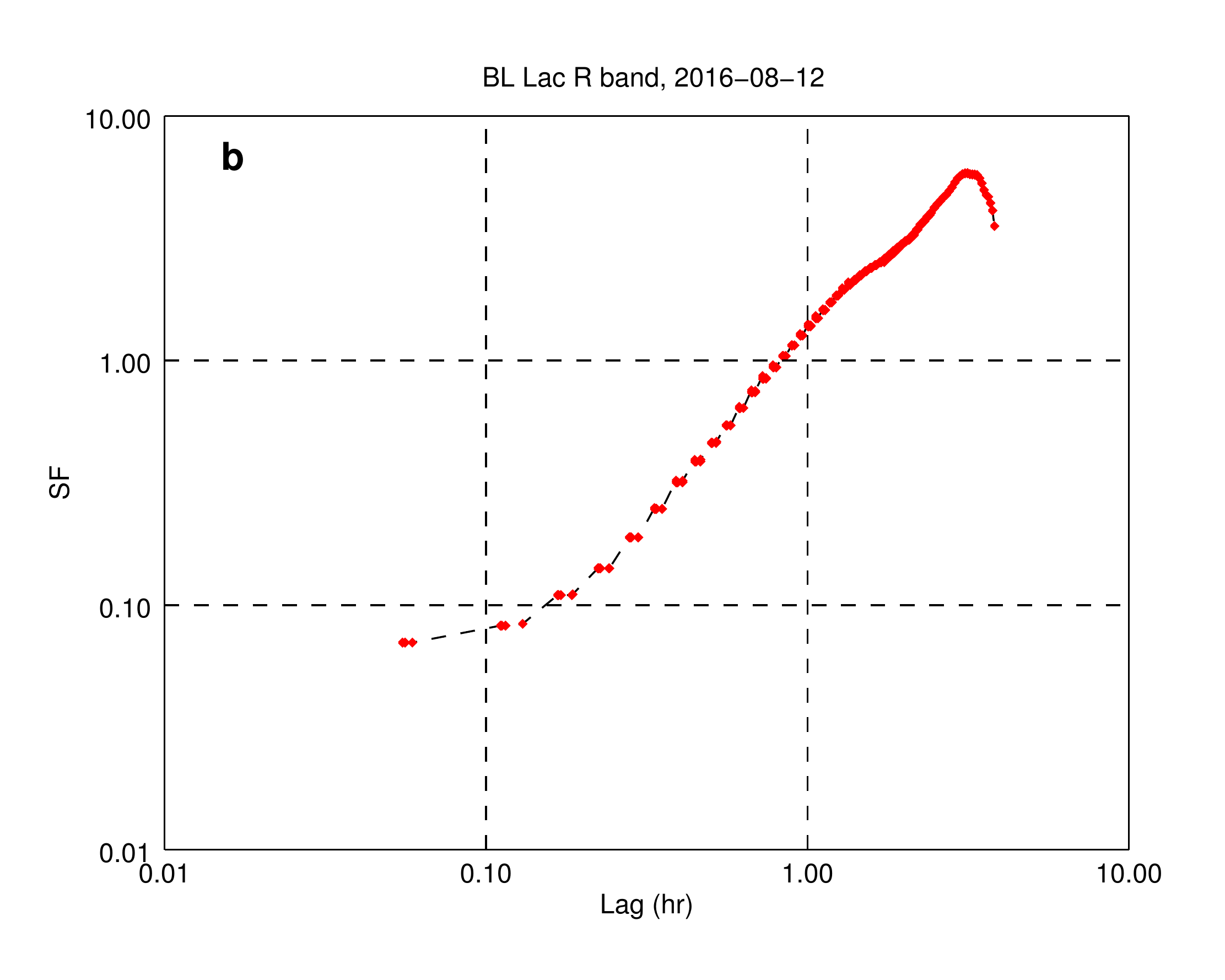}} \\
\resizebox{0.43\textwidth}{!}{\includegraphics{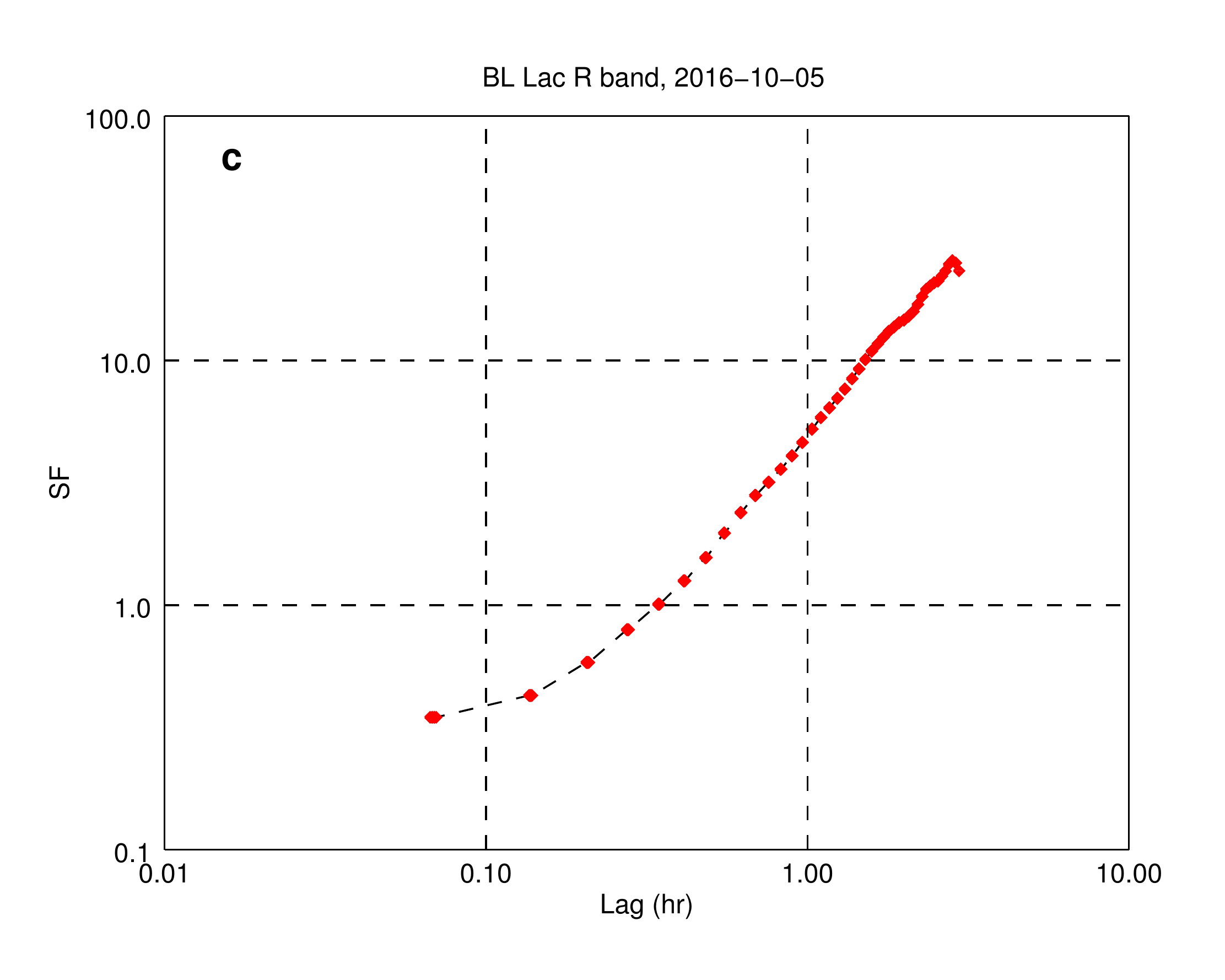}} &
\resizebox{0.43\textwidth}{!}{\includegraphics{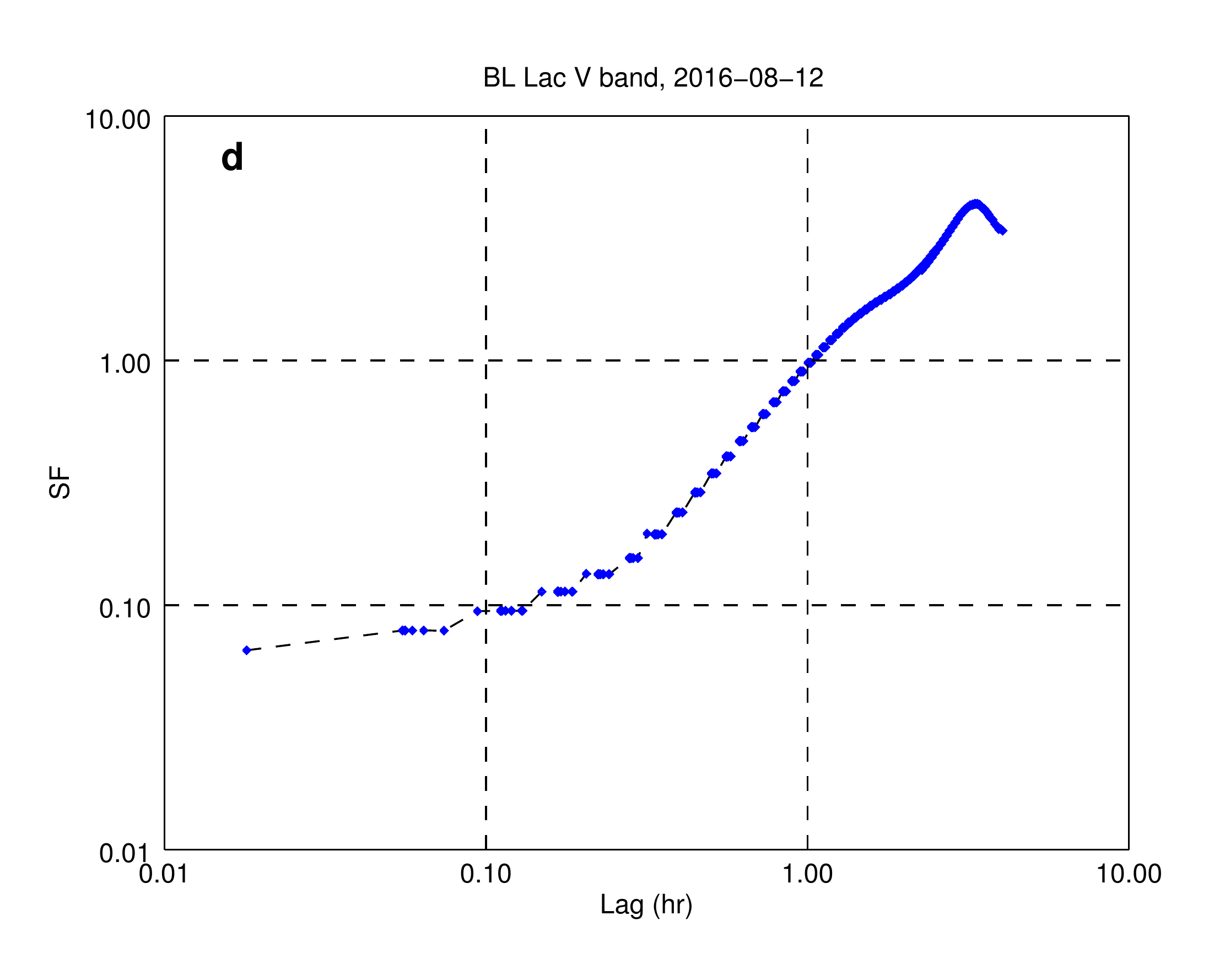}} \\
\resizebox{0.44\textwidth}{!}{\includegraphics{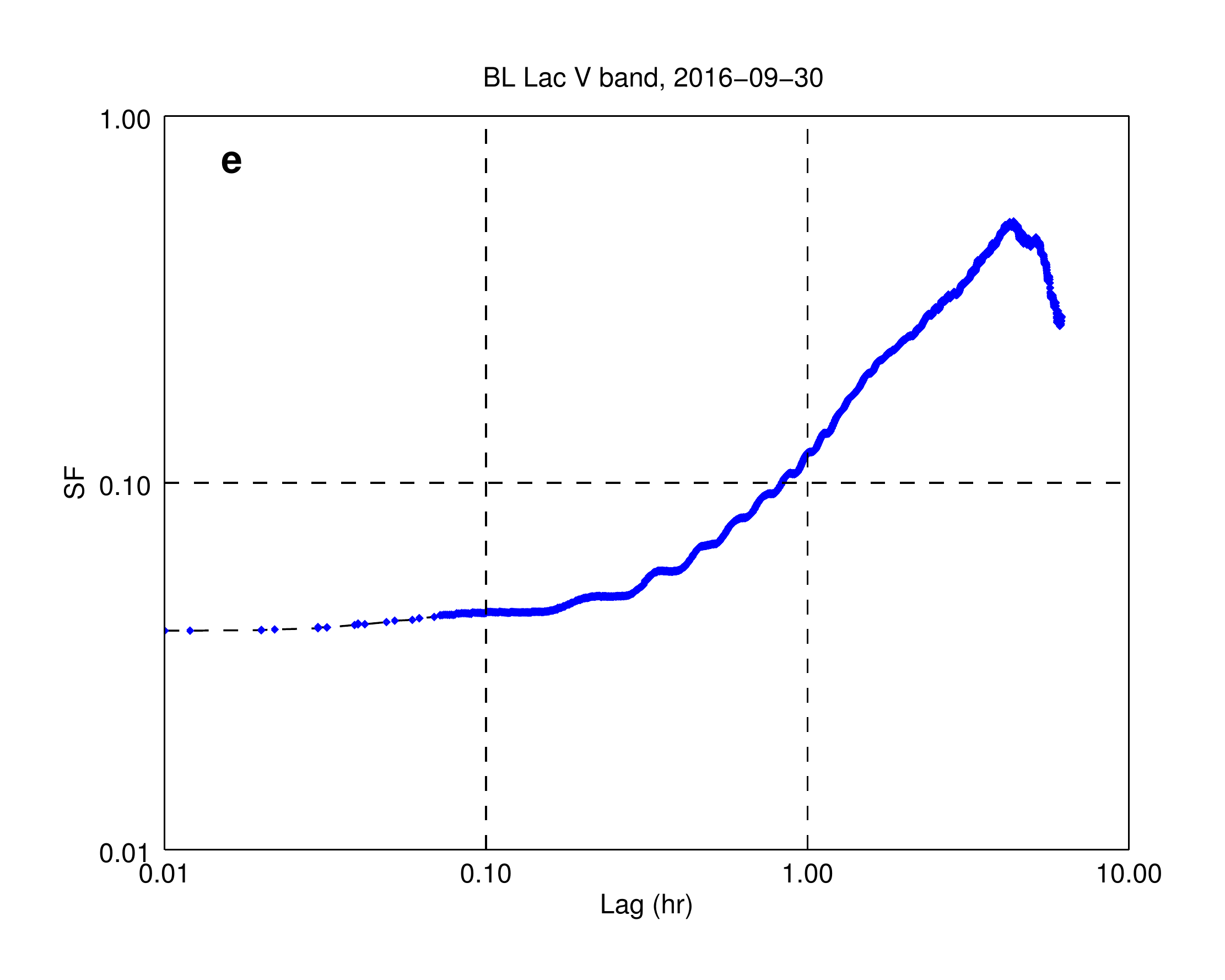}} &
\resizebox{0.44\textwidth}{!}{\includegraphics{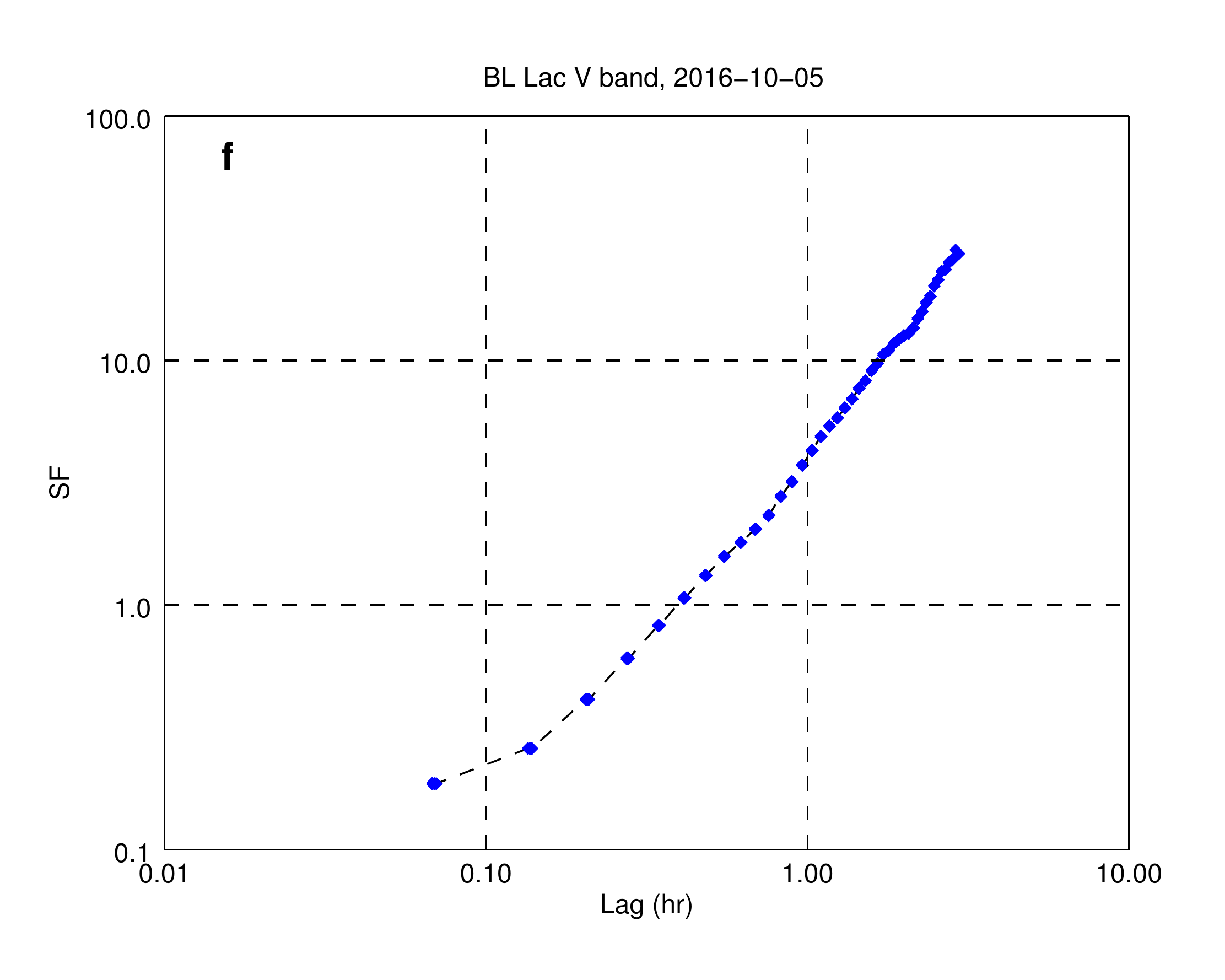}} \\
\resizebox{0.44\textwidth}{!}{\includegraphics{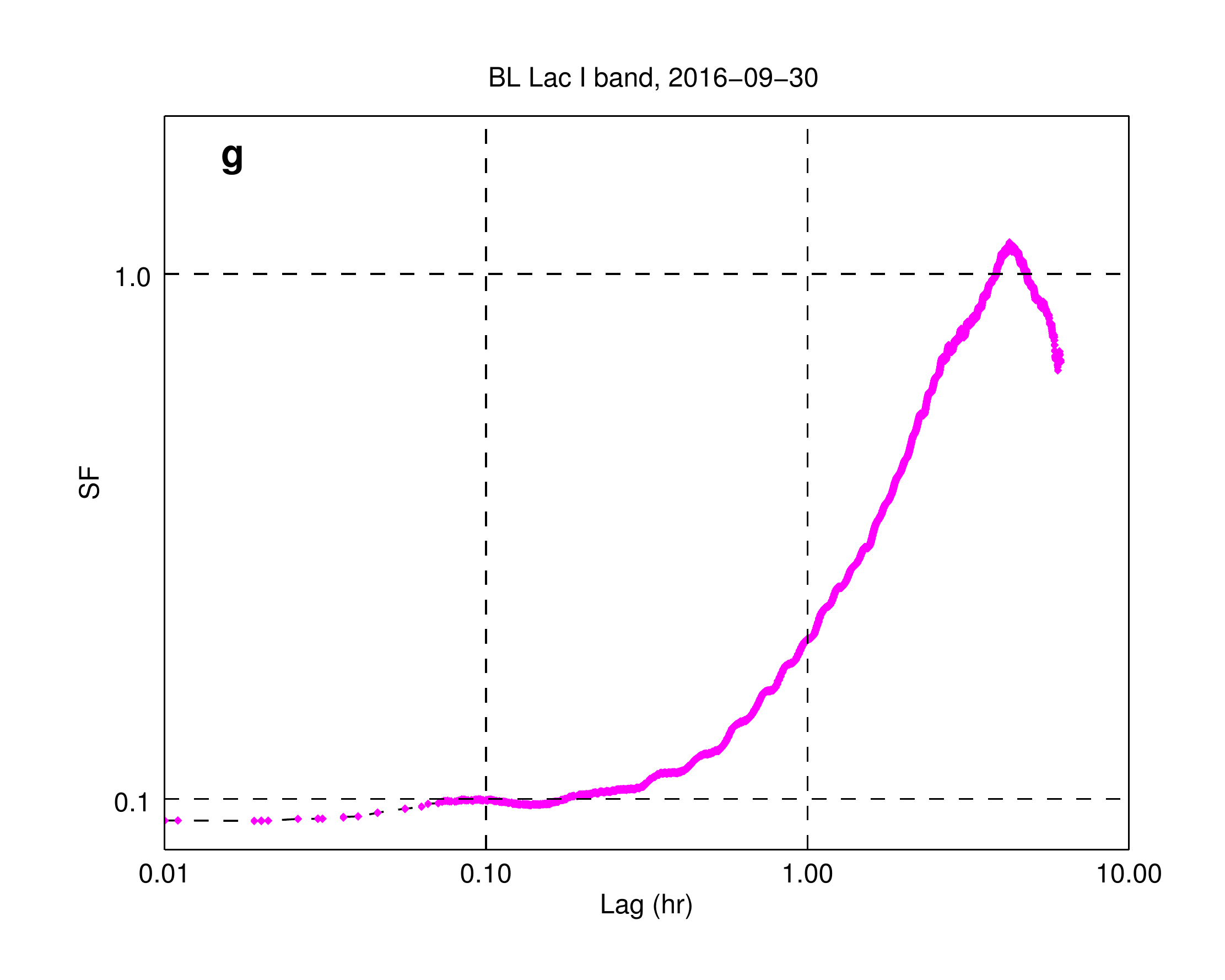}} &
\resizebox{0.44\textwidth}{!}{\includegraphics{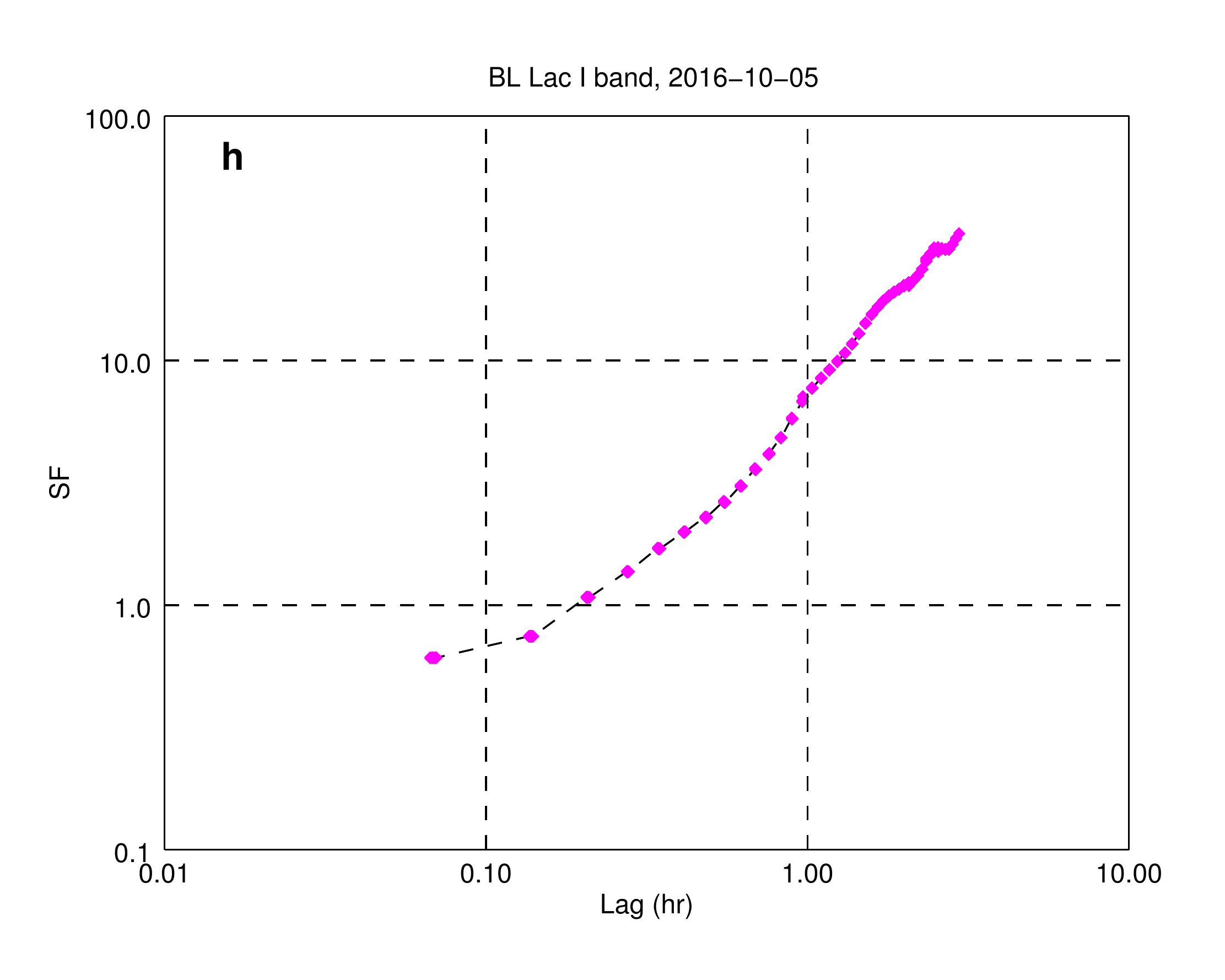}} \\
\end{tabular}
\end{center}
\caption{Intranight structure function of the optical light curves of the blazar BL Lac}
\label{sfun}
\end{figure}

To further investigate the presence of the characteristic timescales,  we also analyzed the multiband observations employing first order structure function \citep{Simonetti85}.  SFs are also widely used in the astronomical time series analysis; and    for a signal x(t) and a signal offset $\tau$, the first order structure function is given by  
 \begin{equation}
 SF\left ( \tau  \right )=\left \langle [x\left ( t \right ) -x\left ( t+\tau  \right )]^{2}\right \rangle
 \label{Sfun}
\end{equation} 
 General interpretations of the SF features are discussed in  \cite{Hughes1992}. The SF can be treated as equivalent to the power spectral density  (PDS) of a light curve calculated in the time domain instead of frequency domain. One of the main advantages of the SF over PSD is that SF method is less sensitive to irregular sampling in the time series and therefore it is relatively free of artifacts similar to windowing and aliasing in frequency domain \cite[e.g.][]{Lainela1993,Paltani1997}. This makes SF one of the favored  tools in time series analysis widely applied in the search for  periodocity' in the AGN light curves \citep[see][and  references therein]{Zhang02} (However for a critical review of SF see  \cite[][]{Emmanoulopoulos10}).  Figure \ref{sfun} shows the structure function plots for all the  microvariability observations of the blazar BL Lac. Although characteristic variability timescales are not obvious in the most of the plots, the SF for R band observations on Night 1 (panel a plot) shows clear break around $\sim0.8$ hr, confirming the presence of such a timescale detected by ACF.

\section{Discussion and Conclusion}
It is seen that the blazar displays multi-band flux variability in the hour-like timescales. Using the causality argument, the minimum variability timescale $\tau_{var}$, can be used to estimate the upper limit of the minimum sizes of the emission region ($R$)  as

\begin{equation}
R=\frac{\delta }{\left ( 1+z \right )}ct_{var}
\label{size}
\end{equation}
where  $\delta$, Doppler factor, is given $\delta =(\Gamma \left ( 1-\beta cos\theta \right ))^{-1}$ and , for the velocity $\beta=v/c$, the bulk Lorentz factor can be written as, $\Gamma=1/\sqrt{1-\beta^{2}}$. In  blazar jets the emission regions  move with high speeds along the path making an angle with the line of sight.  Using a typical variability timescale $t_{var}=1$ hr and source red-shift $z=0.0685$ along with a moderate value of Doppler factor $\delta=10$, the equation \ref{size} can be used to constrain the upper limit of the emission regions size to $\sim1\times10^{15}$\ cm, a size comparable to the solar system. Such small scale fluctuations leading to the rapid intranight multiband variability are most likely intrinsic to the jet  processes and  often hard to be explained  as mere result of extrinsic processes such as geometrical and projection effects often involving variations in Doppler factor of the emission regions. In general, blazar variability should originate within a small, possibly independent sub-volumes of blazar jets, which could be identified with isolated turbulent cells, magnetic reconnection sites,  small-scale shocks induced by turbulence  within the main jet body  \citep[see in this context][]{Narayan2012,bhatta013,Marscher2014,bhatta015,Bhatta2016a}. In particular. Microvariability  displayed by the source might also arise due to magnetohydrodynamical instabilities inducing  localized stochastic particle injections on similar time scales within the turbulent jets of the blazar. Then the combined effects of  the cooling (synchrotron and adiabatic expansion)  on the accelerated particles and the light crossing timescales contribute in shaping the observed fast variability. The frequency dependent variability amplitude i.e. larger variability amplitude  at higher frequency, -- also reported in previous works on intranight variability  \citep[e.g.][]{Nesci1998,Massaro1998,Zhai2012} --  could also be the result of synchrotron emission;  assuming constant magnetic field and isotropic distribution of pitch angles, the synchrotron emission together with $-d\gamma /dt\propto \gamma ^{2}$ and  $t_{syn}\propto 1/\gamma $  could result dissipation larger amount of energy in shorter timescales. For such calculations, BL Lac being a LSP source, it is also considered that  most of the synchrotron radiation is emitted around optical frequency (as characteristic frequency).

 In addition to the rapid flux variability, the observed significant color (spectral) evolution on intranight timescale further help us to delve into the emission regions. In particular, bluer-when-brighter (alternatively harder-when-brighter) trend has been frequently observed in the source during intranight observations \cite[see][]{Zhai2012,Papadakis2003}. Such an achromatic behavior, both bluer-when-brighter and redder-when-brighter, have been frequently observed in the blazars \cite[for further discussion see][]{Bhatta2017,bhatta016}. Similarly, the optical spectra are found to be of the power-law type consistent with the scenario in which the synchrotron emission contribute by power-law particle injections. But Another interesting observation was that the optical spectral revealed a power-spectral break. Such breaks  could arise due to the break in the injected particle distribution. Similar spectral curvatures in the source were also detected during intraday flux changes  in Swift X-ray observations \cite{Raiteri2010}
 and in In hard X-ray (3-79 keV) NuSTAR observation \citep[see][]{Bhatta2017b}.  Such spectral curvature (or the spectral breaks) often indicate the transition region between the low- and high-energy components of the spectral energy distribution (SED) \cite[see][]{Abdo2011}.

The study of the correlation between multiband emission on intranight timescales shows that in most of the cases  both V and R and V and I band emission appear well correlated with each other with no evident lag. This suggests that  microvariability primarily originates from a single emission zone with the similar population of the particles.  But during one of the nights we found that the I band variations lag behind V band variations, also noted in the source by previous authors \cite[e.g.][]{Papadakis2003} on similar timescale. Such lags between multifrequecy emission could be resulted due to frequency dependent opacity in which the higher emission escaped the emission region first followed by the lower energy emission. Alternatively,  the observed lag could be resulted due to the separation of the two (high and low energy) emission regions  \cite[see][]{Fuhrmann2014}. Furthermore, both discrete auto-correlation function and structure function analyses consistently suggest presence of a characteristic timescales of $\sim0.86$ hr in the R band observation of Night 1. However we do not find similar detection in the other observations. This implies the transient nature of processes resulting in the characteristic timescale. Such short timescales characteristic timescales might have resulted due to localized magnetohydrodynamical instabilities either in the accretion disk \cite[e.g.][]{Mangalam1993} or in the relativistic jets \citep[e.g.][]{Hardee1999}. In the disc , such a small timescale processes might occur in the innermost regions whereas in the jets the rest frame timescale might be longer by the Doppler factor of the emission region and thereby leaving door open for a wider range of likely scenarios depending upon the combined effect of the changes in the orientation angle and the bulk speed of the emission region.

\vspace{6pt} 

\acknowledgments{GB acknowledges the support from the Polish National Science Centre grants DEC-2012/04/A/ST9/00083.}

\conflictsofinterest{The authors declare no conflict of interest} 

\abbreviations{The following abbreviations are used in this manuscript:\\

\noindent 
\begin{tabular}{@{}ll}
AGN & Active Galactic Nuclei\\
BL Lac & BL Lacertae object\\
FSRQ & Flat Spectrum Radio Quasar\\
JKT&Jacobus Kapteyn Telescope \\
SARA&Southeastern  Association for Research in Astronomy\\
SED&spectral energy distribution \\
VLBA& Very Long Baseline Array\\
VLBI &Very Long Baseline Interferometry\\
WEBT & Whole Earth Blazar Telescope\\
 VERITAS &Very Energetic Radiation Imaging Telescope Array System

\end{tabular}}


\reftitle{References}



\sampleavailability{Samples of the compounds ...... are available from the authors.}

\end{document}